\documentclass[11pt]{article}
\begin{document}
\newcommand{\eq}[1]{(\ref{#1})}
\def\la{\label}
\def    \beq    {\begin{equation}} \def \eeq    {\end{equation}}
\def    \bea    {\begin{eqnarray}} \def \eea    {\end{eqnarray}}
\def    \lf     {\left (} \def  \rt     {\right )}
\def    \a      {\alpha} \def   \lm     {\lambda}
\def    \D      {\Delta} \def   \r      {\rho}
\def    \th     {\theta} \def   \rg     {\sqrt{g}} \def \Slash  {\, /
\! \! \! \!}  \def      \comma  {\; , \; \;} \def       \pl
{\partial} \def         \del    {\nabla}
\newcommand{\mx}[4]{\left#1\begin{array}{#2}#3\end{array}\right#4}
\newcommand{\od}[1]{$\widetilde{{\rm OD}#1}$}
\newcommand{\ndod}[1]{\widetilde{{\rm OD}#1}}
\newcommand{\gt}[1]{D$#1$-$\widetilde{\rm GT}$}
\newcommand{\ndgt}[1]{{\rm D}#1-\widetilde{\rm GT}}
\rightline{UG-01-26} \rightline{hep-th/0103188}

\frenchspacing \thispagestyle{empty}

\begin{center}\Large \textbf{Open String/Open D-Brane Dualities: Old and New}\\

\end{center}

\vspace{0.5cm}

\begin{center}

\large \textbf{Henric Larsson}$^{1}$ and \textbf{Per Sundell}$^{2}$

\end{center}

\vspace{0.3cm}

\normalsize

\begin{center}

$^{1}$ \emph{Institute for Theoretical Physics \\
G\"oteborg University and Chalmers University of Technology \\
SE-412 96 G\"oteborg, Sweden \\
E-mail: solo@fy.chalmers.se}

\end{center}

\vspace{0.3cm}

\begin{center}

$^{2}$ \emph{Institute for Physics, University of Groningen\\
Nijenborgh 4, 9747 AG Groningen, The Netherlands\\
E-mail: p.sundell@phys.rug.nl}

\end{center}

\vspace{0.8cm}

\begin{center}

\large \textbf{Abstract} \end{center} \normalsize

We examine magnetic and electric near horizon regions of maximally
supersymmetric D-brane and NS5-brane bound states and find
transformations between near horizon regions with worldvolume
dual magnetic and electric fluxes. These point to dual
formulations of NCYM, NCOS and OD$p$ theories in the limit of
weak coupling and large spatial or temporal non-commutativity
length scale in terms of weakly coupled theories with fixed
worldvolume dual non-commutativity based on open D-branes. We
also examine the strong coupling behavior of the open D-brane
theories and propose a unified web of dualities involving
strong/weak coupling as well as large/small non-commutativity
scale.

\newpage

\section{Introduction}

In this paper we examine dualities between decoupled
non-gravitational theories arising on branes in M-theory and
string theory \cite{GMSS}-\cite{Newtonian}. These arise in limits
involving taking the Planck length to zero at fixed energies, or
equivalently, the energy to zero at fixed Planck length, while
keeping a finite effective non-gravitational coupling on the
brane. Starting from a large stack of branes, whose decoupled
dynamics is described by the residual gravitational
M-theory/string theory in its near horizon region, the dynamics
of a single or a few probe branes is obtained by separating them
from the stack and moving them up and away from the gravitational
core of the near horizon region into a UV region where the
gravitational energy scale, set by a running Planck mass,
diverges. This process is of course equivalent to scaling the
Planck length to zero in a flat background, but it has the
advantage of being embedded in a non-perturbative setup. The
decoupling of the probe brane then follows from the decoupling of
the near horizon region from the flat region. Moreover, the
required scalings of tensions and fluxes are easily read off from
the components of the near horizon configuration.

A further advantage is that the underlying U-duality group
descends to the decoupled theories. Discrete elements of the
group (S-dualities, T-dualities and uplifts) generate dualities
between decoupled theories which differ in their perturbative
spectrum.  Continuous elements generate `non-commutative
deformations'. A magnetic deformation affects the geometry of the
worldvolume, but does not alter the perturbative spectrum and
couplings. Electric deformations act differently. A finite
electric field becomes critical in the decoupling limit which
leads to a decoupled theory described in terms of an open brane.
Thus, loosely speaking, there are two kinds of decoupled
non-gravitational theories: `field theories' that arise in
non-critical limits (and their magnetic deformations) and `open
brane theories' that arise in critical limits (and their magnetic
deformations).

M-theory has a critical open-membrane (OM) theory on its
five-brane \cite{GMSS,us2,BermanSundell,harmark2,RussoJabbari},
which subsumes various spatially non-commutative Yang-Mills (NCYM)
theories \cite{con,hull,sw,maldaR,oz,Lu1} and critical
non-commutative open string (NCOS) theories
\cite{gopa,lennat,Harmark1} on the D-branes. While OM theory is
non-perturbative, NCYM and NCOS theories have dimensionless
couplings which are determined from the geometrical data of the
wrapping of the OM theory expressed using the six-dimensional
open membrane metric \cite{us2}. The OM theory also accounts for
a non-commutative extension of the type $(2,0)$ little string
theory (LST) on the IIA NS$5$-brane containing heavy D$2$-branes
called OD$2$ theory \cite{GMSS,harmark2}. T-duality in a
direction parallel to the electric RR flux sends the OD$2$ theory
to a non-commutative extension of the type $(1,1)$ LST on the IIB
NS$5$-brane referred to as the OD$1$ theory \cite{GMSS,harmark2},
that in turn is S-dual to the six-dimensional NCOS theory on the
IIB D$5$-brane. The OD$2$ theory can also be T-dualized along a
magnetic circle to yield OD$3$ theory on the IIB NS$5$ brane
\cite{GMSS,NS5,obers1}. In what follows we shall assume that all
brane directions are non-compact -- compact directions lead to
decoupled theories involving open brane modes as well as wound
closed strings and wrapped higher dimensional branes with
strictly positive winding/wrapping number \cite{Newtonian}.

The above theories are thus related by dualities involving
various limits in the coupling and the non-commutative length
scale. There are, however, several limits in which none of the
above theories provides a perturbative definition. For example,
four-dimensional NCYM theory with large $\theta_{\rm YM}$ and
small $g^2_{\rm YM}$ and fixed NCYM soliton tension $(g_{\rm
YM}^2\theta_{\rm YM})^{-1}$, or equivalently four-dimensional
NCOS with fixed $\alpha'_{\rm eff}$ and large $G^2_{\rm OS}$,
should be dual to a theory with a well-defined expansion in
$g^2_{\rm YM}=1/G^2_{\rm OS}$ at fixed $\alpha'_{\rm
eff}=\theta_{\rm YM}g^2_{\rm YM}$. There are at least four
possibilities for what interactions this theory may have. They
may be the ones of the large $\theta_{\rm YM}$ and small
$g^2_{\rm YM}$ limit of NCYM theory. It may also have weak
self-interactions of NCOS type provided this theory is actually
self-dual, though this seems artificial from the underlying IIB
theory perspective as its self-duality simply maps the NCYM
supergravity dual to a strongly coupled NCOS dual. The theory may
also be free.

A fourth option is that the theory has non-trivial interactions
different from both NCOS and NCYM theory. The supergravity dual
suggests interactions based on critical open D-strings. From the
NCYM viewpoint a finite $\theta_{\rm YM}$ tilts an open D-string
a finite angle \cite{hashimoto,gross1}. Such a sub-critical open
D-string casts a `shadow' on the D$3$-brane giving rise to a
non-local NCYM soliton. The F-strings of the NCOS theory become
critical in a discontinuous way whereby the decoupled open
F-string `dipoles' arise only in the critical limit whereas any
sub-critical electric field leads to essentially the same coupled
open-closed string dynamics. Similarly, the critical transition
of the open D-string arises in the limit of infinite $\theta_{\rm
YM}$ when it coincides with its shadow. Then new massless states
associated with open F-strings of zero length stretched between
the open D-string and the D$3$-brane appears in the spectrum.
Therefore the usage of the term `critical open D-brane' requires
extra care when meant to suggest an analogy with critical open
F-strings. In this paper we shall not examine the microscopic
aspects of this issue any further. We shall adopt the point of
view, however, that there exists some non-trivial theory, which
we shall denote by \od1, that has a well-defined perturbative
expansion in $G^2_{\ndod1}=g^2_{\rm YM}$ at fixed
$\ell^2_{\ndod1}=\theta_{\rm YM}g^2_{\rm YM}$ (thus involving
critical open D-strings) and that has a supergravity dual
obtained by taking the electric near horizon limit of a
D$3$-brane with electrically deformed RR two-form potential (see
below for details). We also wish to remark\footnote{We are
thankful to A. Hashimoto for pointing this out to us.} that from
the macroscopic point of view the usual two-dimensional phase
diagram in the energy-coupling plane does not contain the new
\od1 phase, as this diagram contains the phases of the theory
expanded around its usual vacuum with zero number of
non-commutative solitons. It is expected, however, that the \od1
theory arises as a stable thermodynamic phase provided a chemical
potential for the solitons is added, such that the theory can be
examined in an environment with finite expectation value for the
soliton number.

Having made this assumption, it is natural to extend the notion
of open D$q$-brane theories in $(q+3)$ dimensions to $q=0,1,2,3$.
We shall thus discuss \od{q} theories based on D$(q+2)$-brane
supergravity duals with electric RR $(q+1)$-form flux and that we
assume have a well-defined perturbative expansion in a coupling,
to be specified below, at fixed temporal non-commutativity equal
to a critical open D$q$-brane tension (see above remark). We
shall also discuss generalized gauge theories, which we shall
refer to as D$q$-GT, based on D$(q+2)$-brane supergravity duals
with magnetic RR $(q+1)$-form flux and that we assume have a
well-defined perturbative expansion at fixed spatial
non-commutativity.  The D$q$-GT also have analogs living on
NS$5$-branes, which we shall refer to as \gt{q}, and that are
generalized gauge theories in the IIB theory and tensor
counterparts in IIA. In fact, the \gt{q} are magnetic
deformations of the tensionless little string theories on the
type IIA/B NS$5$-branes \cite{20}, which may serve as a separate
motivation for including these theories. Though we shall base all
our observations on the D$q$-GT and \gt{q} on supergravity duals,
we wish to think of them as zero-slope limits of open D-brane
theories similar to those yielding NCYM theories from open
strings.

In the remainder of this paper we shall examine the supergravity
duals of these theories and deduce an extended and unified
duality web which incorporates limits of large couplings as well
as large spatial and temporal non-commutativity length scales.

The plan of the paper is as follows: In Section 2 we give the
supergravity solutions with special attention paid to the
different parameterizations resulting from electric and magnetic
deformations. In Section 3 we show the equivalence between
electric and dual magnetic near horizon regions. The new open
D-brane theories are examined in Section 4 and 5. Section 6
contains a summary and discussion. The conventions used for
S-duality, T-duality and uplifting to 11-dimensions are given in
an Appendix.

\section{The supergravity solutions}

In this section we obtain each of the half-supersymmetric
D$p$-F$1$, D$(q+2)$-D$q$ and NS$5$-D$p$ bound states by either a
magnetic or a worldvolume-dual electric deformation.  We then
obtain the supergravity duals by taking magnetic or electric near
horizon limits, respectively, after which we define couplings and
length scales. These identifications shall be discussed in more
detail in Sections 4 and 5.

\subsection{NS-deformations of D$p$-branes (NCOS and NCYM)}

We start by obtaining the D$p$-F1 and D$p$-D($p-2$) bound states
by deforming a half-supersymmetric D$p$-brane with an electric
$B$-field and a magnetic $B$-field, respectively. The resulting
electric bound states are given by\footnote{There are many ways of
constructing these solutions. For a general method based on a
parametrization directly in terms of the $O(p+1,p+1)$ generators
we refer to \cite{Per,solo}. For other methods of constructing
bound states, see e.g.
\cite{townsend,Russo,solutions,Costa,martin,Lu2}. }

\bea ds^2&=&H^{-{1\over 2}}\left({1\over
h_-}(-dx_0^2+dx_1^2)+dx_2^2 +\ldots+dx_p^2\right)+H^{1\over
2}(dr^2+r^2d\Omega^2_{8-p})\
,\nonumber\\ e^{2\phi}&=&{1\over h_-}g^2H^{\frac{3-p}2}\ , \nonumber \\
B&=&{\theta \over\alpha'Hh_-}dx^0\wedge dx^1 \ ,\label{Epbrane}
  \\
C&=&\frac1{gHh_-}dx^0\wedge \cdots\wedge
dx^p+(7-p)N(\alpha')^{\frac{7-p}2}(1-B)\wedge \epsilon_{7-p}\nonumber \\
&&-{\theta\over gH\alpha'}dx^2\wedge\ldots\wedge dx^p\ ,
\nonumber\\
\mbox{where}&& H=1+ \frac{gN(\alpha')^{\frac{7-p}{2}}}{r^{7-p}}\
,\nonumber \eea and the magnetic ones by

\bea ds^2&=&H^{-{1\over 2}}\left(-dx_0^2+\ldots+dx^{2}_{p-2}+
\frac{1}{h_+}(dx_{p-1}^{2}+dx^{2}_{p})\right)+H^{1\over 2}(dr^2+
r^2d\Omega^2_{8-p})\,\nonumber\\ e^{2\phi}&=&{1\over
h_+}g^2H^{\frac{3-p}2}\ ,
 \nonumber \\
B&=&-{\theta \over\alpha'Hh_+}dx^{p-1}\wedge dx^{p} \
,\label{Mpbrane}
  \\
C&=&\frac1{gHh_+}dx^0\wedge \cdots\wedge
dx^p+(7-p)N(\alpha')^{\frac{7-p}2}(1-B)\wedge \epsilon_{7-p}\nonumber \\
&&-{\theta\over gH\alpha'}dx^0\wedge dx^1\wedge\ldots\wedge dx^{p-2}\ ,
\nonumber\\
\mbox{where}&&
 H=1+\frac{gN(\alpha')^{\frac{7-p}{2}}}{r^{7-p}}\ .
\nonumber \eea Here $d\epsilon_{7-p}$ is the volume element on
the transverse $(8-p)$-sphere and we have defined:

\beq h_\pm\equiv1\pm\left({\theta\over\alpha'}\right)^2H^{-1}\
.\eeq The supergravity duals of the NCOS and NCYM theories are
near-horizon regions of these solutions where $ds^2/\alpha'$,
$B/\alpha'$ and $C_{(k)}/(\alpha')^{k+1\over 2}$ ($k=0,1,2,...$)
are held fixed in the limit $\alpha' \rightarrow 0$. This can be
achieved by keeping the following quantities fixed\footnote{In
the electric case \eq{osnh} leads to that $\alpha'$ cancels out,
so that the limit $\alpha'\rightarrow 0$ becomes trivial. The
electric limit is therefore equivalent to setting
$\tilde{\ell}=\sqrt{\alpha'}$, and keeping $\tilde{\ell}$, $g$
and all coordinates fixed while taking $\theta\rightarrow
\alpha'$. In the magnetic case, we keep $g$ fixed in (\ref{ymnh})
when $p=3$. This means that $\ell^{p-3}$ or $\ell^{q-1}$ should
be changed to $g$ when $p=q+2=3$ in (\ref{NCYMbrane}) and
(\ref{Dymbrane}).}:

\bea \mbox{Electric near-horizon}&:& \quad
\tilde{x}^{\mu}={\tilde{\ell} x^{\mu}\over \sqrt{\alpha'}},\
\tilde{r}={\tilde{\ell} r\over\sqrt{\alpha'}},\
{\theta\over\alpha'}=1,\ g,\ \tilde{\ell}\
.\label{osnh}\\\mbox{Magnetic near-horizon}&:& \quad x^\mu,\
u=\frac{r}{\alpha'},\ \theta,\nonumber\\&&
\left\{\begin{array}{ll}\ell^{p-3}= g(\alpha')^{\frac{p-3}2}&p\neq
3\\ g&p=3\end{array}\right.\ .\label{ymnh}\eea Here $\ell$ and
$\tilde{\ell}$ are fixed length scales. The electric near horizon
region of (\ref{Epbrane}) gives the supergravity dual of the
$(p+1)$-dimensional NCOS theory:

\bea \frac{ds^{2}}{\alpha'}&=&\frac{1}{\tilde{\ell}^2}H^{-{1\over
2}}\Big[H\Big( \frac{\tilde{r}}{\tilde{R}}\Big)
^{7-p}(-d\tilde{x}_{0}^{2}+d\tilde{x}^{2}_{1})+d\tilde{x}^{2}_{2}+\ldots
 +d\tilde{x}_{p}^{2} \nonumber\\
&+&H(d\tilde{r}^{2}+\tilde{r}^{2}d\Omega^{2}_{8-p})\Big]\ ,
\nonumber\\
e^{2\phi}&=&g^{2}H^{\frac{5-p}{2}}\Big(\frac{\tilde{r}}{\tilde{R}}
\Big)^{7-p}\ , \nonumber\\ \frac{C_{01\ldots
p}}{(\alpha')^{\frac{p+1}{2}}}&=&\frac{1}{g\tilde{\ell}^{p+1} }
\Big(\frac{\tilde{r}}{\tilde{R}}\Big)^{7-p}\ ,\nonumber\\
\frac{C_{23\ldots
p}}{(\alpha')^{\frac{p-1}{2}}}&=&-\frac{1}{g\tilde{\ell}^{p-1}}
\frac{1}{H}\ ,
 \label{NCOSbrane} \\
\frac{B_{01}}{\alpha'}&=&\frac{1}{\tilde{\ell}^2}\Big(\frac{\tilde{r}}
{\tilde{R}}\Big)^{7-p}\ , \nonumber\\ \mbox{where}&&
H=1+\Big(\frac{\tilde{R}}{\tilde{r}}\Big)^{7-p}\ ,\quad
\tilde{R}^{7-p}=Ng\tilde{\ell}^{7-p} \ .\nonumber \eea The NCOS
length scale and coupling are given by

\beq \alpha'_{\rm eff}=\tilde{\ell}^2\ ,\quad G^2_{\rm OS}=g\ .
\la{NCOSp}\eeq
The magnetic near horizon region of (\ref{Mpbrane}) gives the
supergravity dual of the $(p+1)$-dimensional NCYM theory ($p\neq
3$; for $p=3$ we replace $\ell^{p-3}$ by $g$):

\bea
\frac{ds^{2}}{\alpha'}&=&\Big(\frac{u}{R}\Big)^{\frac{7-p}{2}}\Big(-dx_{0}^{2}
+\ldots +dx_{p-2}^{2}+\frac{1}{h} (dx_{p-1}^{2}+dx_{p}^{2})\Big)
\nonumber\\
&+&\Big(\frac{R}{u}\Big)^{\frac{7-p}{2}}(du^{2}+u^{2}d\Omega^{2}_{8-p})
\ , \nonumber\\
e^{2\phi}&=&\ell^{2p-6}h^{-1}\Big(\frac{R}{u}\Big)^{{(7-p)(3-p)\over
2}}\ , \nonumber\\ \frac{C_{01\ldots
p}}{(\alpha')^{\frac{p+1}{2}}}&=&\frac{1}{\ell^{p-3}h}
\Big(\frac{u}{R}\Big)^{7-p}\ ,\nonumber\\ \frac{C_{01\ldots
p-2}}{(\alpha')^{\frac{p-1}{2}}}&=&-\frac{\theta}{\ell^{p-3}}
\Big(\frac{u}{R}\Big)^{7-p}\ ,
 \label{NCYMbrane} \\
\frac{B_{p-1,p}}{\alpha'}&=&-\frac{\theta}{h}\Big(\frac{u}
{R}\Big)^{7-p} \ , \nonumber\\ \mbox{where}&&
h=1+\theta^{2}\Big(\frac{u}{R}\Big)^{7-p}\ ,\quad
R^{7-p}=\ell^{p-3}N \ .\nonumber \eea
The Yang-Mills coupling and non-commutativity parameter are
identified as

\beq g_{\rm YM}^2=\mx{\{}{ll}{\ell^{p-3}&p\neq 3\\g&p=3}{.}\
,\quad \theta_{\rm YM}=\theta\ . \la{NCYMp}\eeq We remark that the
non-renormalizability of NCYM theory for $p>3$ corresponds to
large open string coupling in the UV, signaling large open string
fluctuations. For $p=3$ the open string coupling remains finite
and equal to $g_{\rm YM}^2$, while the tension still diverges so
that the massive open string modes decouple. For $p=2$ the open
string coupling goes to zero in the UV. Instead it diverges in
the IR, so that the interacting three-dimensional Yang-Mills
theory is defined in the IR-limit.

\subsection{Electric RR-deformation of NS$5$-branes (OD$p$)}

The half-supersymmetric NS$5$-D$p$ bound states can be obtained
by first S-dualizing the D$5$-F$1$ solution (\ref{Epbrane}) and
then applying a series of T-dualities. This is equivalent to
starting from a NS$5$-brane and turning on an electric
$(p+1)$-form RR-potential $C_{01\ldots p}$. After
rescaling\footnote{The rescaling of the coordinates is part of the
solution generation. However, in the identification of two
supergravity duals, it is more natural to keep the coordinates
fixed and instead rescale the worldvolume parameters; see also
comment in Section 3.} $x^{\mu} \rightarrow\sqrt{g}x^{\mu}$ and
$r\rightarrow \sqrt{g}r$ and defining $\tilde{g}=g^{-1}$ we find:

\bea ds^2&=&h_-^{1\over 2}\Big[{1\over h_-}(-dx_0^2+\ldots
+dx_{p}^{2})+dx_{p+1}^{2}+\ldots
 +dx_{5}^{2} \nonumber\\
&+&H(dr^{2}+r^{2}d\Omega^{2}_{3})\Big)]\ ,\nonumber\\
e^{2\phi}&=&\tilde{g}^{2}Hh_-^{\frac{3-p}{2}}\ ,\nonumber\\
C_{01\ldots p}&=&-{\theta\over\alpha'\tilde{g}Hh_-} \ , \label{Npbrane} \\
C_{p+1\ldots 5}&=&-\frac{\theta}{\alpha'\tilde{g}H}\ ,\nonumber\\
B&=&\alpha' 2N\epsilon_{2}\nonumber\\
\mbox{where}&&
 H=1+\frac{N\alpha'}{r^{2}}.\nonumber
\eea
Taking the electric near horizon limit (\ref{osnh}) gives
the supergravity duals of the ODp theories:

\bea \frac{ds^{2}}{\alpha'}&=&\frac{1}{\tilde{\ell}^2H^{1\over 2
}}\frac{\tilde{R}}
{\tilde{r}}\Big[H\left(\frac{\tilde{r}}{\tilde{R}}\right)^2(-d\tilde{x}_{0}^{2}+\ldots
 +d\tilde{x}_{p}^{2})+d\tilde{x}_{p+1}^{2}+\ldots
 +d\tilde{x}_{5}^{2} \nonumber\\
&+&H(d\tilde{r}^{2}+\tilde{r}^{2}d\Omega^{2}_{3})\Big]\ ,
\nonumber\\
e^{2\phi}&=&\tilde{g}^{2}H^{\frac{p-1}{2}}\Big(\frac{\tilde{r}}{\tilde{R}}
\Big)^{p-3}\ , \nonumber\\ \frac{C_{01\ldots
p}}{(\alpha')^{\frac{p+1}{2}}}&=&-\frac{1}{\tilde{\ell}
^{p+1}\tilde{g}}\left(\frac{\tilde{r}}{\tilde{R}}\right)^2\ ,
 \label{ODpbrane} \\
\frac{C_{p+1\ldots
5}}{(\alpha')^{\frac{5-p}{2}}}&=&-\frac{1}{\tilde{\ell}
^{5-p}\tilde{g}H}\ , \nonumber\\
\frac{B}{\alpha'}&=&\frac{2\tilde{R}^{2}\epsilon_{2}}{\tilde{\ell}^2}
\nonumber\\ \mbox{where}&&
H=1+\frac{\tilde{R}^{2}}{\tilde{r}^{2}}\ ,\quad
\tilde{R}^{2}=N\tilde{\ell}^2 \ .\nonumber \eea The little string
length and OD$p$ length and coupling are defined as
\cite{GMSS,harmark2}:

\beq \ell_{\rm LST}=\tilde{\ell}\ ,\quad \ell_{\rm
ODp}=\tilde{g}^{1\over p+1}\tilde{\ell}\ ,\quad G^2_{{\rm
OD}p}=\tilde{g}\ .\la{ODpp}\eeq

\subsection{Electric RR-deformation of D-branes (\od{q})}

Half-supersymmetric D$(q+2)$-D$q$ bound states can be obtained by
S-dualizing the D$3$-F$1$ bound state (\ref{Epbrane}) and then
T-dualizing (together with transverse smearings/localizations).
This procedure is equivalent to starting with a D$(q+2)$-brane
and turning on an electric $(q+1)$-form RR-potential $C_{01\ldots
q}$.  After rescaling $x^{\mu}\rightarrow \sqrt{g}x^{\mu}$,
$r\rightarrow \sqrt{g}r$ and sending $g\rightarrow 1/g$ we find:

\bea ds^2&=&h_-^{1\over 2}\Big[H^{-{1\over 2}}\Big({1\over
h_-}(-dx_0^2+\ldots +dx_{q}^{2})
+dx_{q+1}^{2}+dx_{q+2}^{2}\Big) \nonumber\\
&+&H^{1\over 2}(dr^{2}+r^{2}d\Omega^{2}_{6-q})\Big]\ ,\nonumber\\
e^{2\phi}&=&g^{2}H^{\frac{1-q}{2}}h_-^{\frac{3-q}{2}}\ , \nonumber\\
C_{01\ldots q+2}&=&\frac{1}{gH}\ ,\nonumber\\
C_{01\ldots q}&=&-{\theta\over\alpha'gHh_-} \ , \label{Dpbrane} \\
B_{q+1,q+2}&=&-\frac{\theta}{\alpha'H}\ ,\nonumber\\
\mbox{where}&&
 H=1+\frac{gN\alpha'^{\frac{5-q}{2}}}{r^{5-q}}\ .\nonumber
\eea The electrical near horizon limit ({\ref{osnh}) yields the
following configuration, which we shall interpret as the
supergravity dual of the \od{q} theory:

\bea \frac{ds^{2}}{\alpha'}&=&\frac{1}{\tilde{\ell}^2 H
}\Big(\frac{\tilde{R}}
{\tilde{r}}\Big)^{\frac{5-q}{2}}\Big[H\left(\frac{\tilde{r}}{\tilde{R}}\right)
^{5-q}(-d\tilde{x}_{0}^{2}+\ldots
 +d\tilde{x}_{q}^{2})+d\tilde{x}_{q+1}^{2}+d\tilde{x}_{q+2}^{2} \nonumber\\
&+&H(d\tilde{r}^{2}+\tilde{r}^{2}d\Omega^{2}_{6-q})\Big]\ ,
\nonumber\\
e^{2\phi}&=&\frac{g^{2}}{H}\Big(\frac{\tilde{R}}{\tilde{r}}
\Big)^{{(5-q)(3-q)\over 2}}\ , \nonumber\\ \frac{C_{01\ldots
q+2}}{(\alpha')^{\frac{q+3}{2}}}&=&\frac{1}{g\tilde{\ell} ^{q+3}H}\
,\nonumber\\ \frac{C_{01\ldots
q}}{(\alpha')^{\frac{q+1}{2}}}&=&-\frac{1}{g\tilde{\ell}^{q+1}}
\Big(\frac{\tilde{r}}{\tilde{R}}\Big)^{5-q}\ ,
 \label{ODqbrane} \\
\frac{B_{q+1,q+2}}{\alpha'}&=&-\frac{1}{\tilde{\ell} ^2H} \ ,
\nonumber\\ \mbox{where}&&
H=1+\frac{\tilde{R}^{5-q}}{\tilde{r}^{5-q}}\ ,\quad
\tilde{R}^{5-q}=Ng\tilde{\ell}^{5-q} \ .\nonumber \eea We
identify the \od{q} length and coupling as follows

\beq \ell_{\ndod{q}}=g^{1\over q+1}\tilde{\ell}\ ,\quad
G^2_{\ndod{q}}=g\ .\la{Odqp}\eeq

\subsection{Magnetic RR-deformation of D-branes (D$q$-GT)}

In this section we deform D$(q+2)$-branes with magnetic RR
$(q+1)$-form potentials. In Section 5.1 we shall argue that the
magnetic near horizon region of the resulting half-supersymmetric
D$(q+2)$-F$1$ bound state describes a generalized non-commutative
gauge theory. The bound states can be obtained by S-dualizing a
magnetically (NS) deformed D$3$-brane and then T-dualizing
(together with smearings/localizations). After rescaling
$x^{\mu}\rightarrow \sqrt{g}x^{\mu}$, $r\rightarrow \sqrt{g}r$
and sending $g\rightarrow 1/g$ we find:

\bea ds^2&=&h_+^{1\over 2}\Big[H^{-{1\over 2}
}\Big(-dx_0^2+dx_{1}^{2}
+\frac{1}{h_+}(dx_{2}^{2}+\ldots +dx_{q+2}^{2})\Big) \nonumber\\
&+&H^{1\over 2}(dr^{2}+r^{2}d\Omega^{2}_{6-q})\Big]\ ,\nonumber\\
e^{2\phi}&=&g^{2}H^{\frac{1-q}{2}}h_+^{\frac{3-q}{2}}\ , \nonumber\\
C_{01\ldots q+2}&=&\frac{1}{gH}\ ,\nonumber\\
C_{2\ldots q+2}&=&{\theta\over\alpha'gHh_+} \ , \label{Dpmbrane} \\
B_{01}&=&-\frac{\theta}{\alpha'H}\ ,\nonumber\\
\mbox{where}&&
 H=1+\frac{gN\alpha'^{\frac{5-q}{2}}}{r^{5-q}}\ .\nonumber
\eea Taking the magnetic near horizon limit (\ref{ymnh}) gives
the D$q$-GT dual ($q\neq 1$; for $q=1$ we replace $\ell^{q-1}$ by
$g$):

\bea \frac{ds^{2}}{\alpha'}&=&h^{1\over 2}\Big[\Big(\frac{u}
{R}\Big)^{\frac{5-q}{2}}\Big(-dx_{0}^{2}+dx_{1}^{2}+\frac{1}{h}
(dx_{2}^{2}+\ldots+dx_{q+2}^{2})\Big) \nonumber\\
&+&\Big(\frac{R}{u}\Big)^{\frac{5-q}{2}}(du^{2}+u^{2}d\Omega^{2}_{6-q}
)\Big]\ , \nonumber\\
e^{2\phi}&=&\ell^{2(q-1)}h^{\frac{3-q}{2}}\Big(\frac{R}{u}
\Big)^{{(5-q)(1-q)\over 2}}\ , \nonumber\\ \frac{C_{01\ldots
q+2}}{(\alpha')^{\frac{q+3}{2}}}&=&\ell^{1-q}
\Big(\frac{u}{R}\Big)^{5-q}\ ,\nonumber\\ \frac{C_{2\ldots
q+2}}{(\alpha')^{\frac{q+1}{2}}}&=&\frac{\ell^{1-q}\theta}{h}
\Big(\frac{u}{R}\Big)^{5-q}\ ,
 \label{Dymbrane} \\
\frac{B_{01}}{\alpha'}&=&-\theta\Big(\frac{u}{R}\Big)^{5-q} \ ,
\nonumber\\ \mbox{where}&&
h=1+\theta^{2}\Big(\frac{u}{R}\Big)^{5-q}\ ,\quad
R^{5-q}=N\ell^{q-1} \ .\nonumber \eea We identify the D$q$-GT
coupling and non-commutativity parameter as follows:

\beq g_{{\rm D}q{\rm -GT}}^2=\mx{\{}{ll}{
\ell^{q-1}&q\neq1\\g&q=1}{.}\ ,\quad \theta_{{\rm D}q{\rm -GT}} =
\ell^{q-1}\theta\ .\la{DqGTp}\eeq

\subsection{Magnetic RR-deformation of NS$5$-branes (\gt{q})}

The generalized gauge theories on NS$5$-branes, which we shall
denote by \gt{q} ($q$ even for type IIA and $q$ odd for type IIB),
arise in the near horizon region of an NS$5$-brane with a
magnetic $(q+1)$-form RR potential that is fixed in the UV. These
solutions are constructed by S-dualizing (\ref{Dpmbrane}) for
$p=5$ and then T-dualizing, which gives the following
NS$5$-D$(4-q)$ bound states after the rescalings
$x^{\mu}\rightarrow \sqrt{g}x^{\mu}$ and $r\rightarrow \sqrt{g}r$,
$\theta \rightarrow -\theta$ and setting $g\rightarrow 1/g$:

\bea ds^2&=&h_+^{{1\over 2}}\Big(-dx_0^2+\ldots +dx_{4-q}^{2}
+\frac{1}{h_+}(dx_{5-q}^{2}+\ldots +dx_{5}^{2}) \nonumber\\
&+&H(dr^{2}+r^{2}d\Omega^{2}_{3})\Big)\ ,\nonumber\\
e^{2\phi}&=&g^{2}Hh_+^{\frac{3-q}{2}}\ , \nonumber\\
C_{01\ldots 4-q}&=&-\frac{\theta}{\alpha'gH}\ ,\nonumber\\
C_{5-q\ldots 5}&=&-{\theta\over\alpha'gHh_+} \ , \label{NSmbrane} \\
B&=&\alpha'2N\epsilon_{2}\ ,\nonumber\\
\mbox{where}&&
 H=1+\frac{N\alpha'}{r^2}\ .\nonumber
\eea For $p=5$ the magnetic near horizon limit (\ref{ymnh}) on
the D$5$-brane amounts to keeping $g_{YM}^{2}=\alpha'g$ fixed,
which in turn implies that $g\sim \alpha'^{-1}$. The manipulations
used in obtaining \eq{NSmbrane} imply that the magnetic
near-horizon limit of \eq{NSmbrane} amounts to taking
$\alpha'\rightarrow 0$ keeping the following quantities fixed:

\begin{equation} \mbox{Magnetic NS$5$ near-horizon}: \quad
\hat{x}^{\mu}=\frac{\hat{\ell}x^\mu} {\sqrt{\alpha'}},\
\hat{u}=\frac{\hat{\ell}r}{(\alpha')^{3/2}},\ \theta,\
\hat{\ell}^2= g^{-1}\alpha'\ .\label{ymnh2} \end{equation} The
resulting magnetic near-horizon region becomes

\bea
\frac{ds^{2}}{\alpha'}&=&{h^{1\over 2}\over
 \hat{\ell}^2}\Big(-d\hat{x}_{0}^{2}+\ldots +
d\hat{x}_{4-q}^{2}+\frac{1}{h} (d\hat{x}_{5-q}^{2}+\ldots
+d\hat{x}_{5}^{2}) \nonumber\\
&+&\Big(\frac{\hat{R}}{\hat{u}}\Big)^{2}(d\hat{u}^{2}+\hat{u}^{2}d\Omega^{2}_{3})
\Big)\ , \nonumber\\ e^{2\phi}&=&{h^{\frac{3-q}{2}}\over
 \hat{\ell}^{4}}\Big(\frac{\hat{R}}{\hat{u}}\Big)^{2}\ , \nonumber\\
\frac{C_{01\ldots 4-q}}{(\alpha')^{\frac{5-q}{2}}}&=&-\theta
\hat{\ell}^{q-3}\Big(\frac{\hat{u}}{\hat{R}}\Big)^{2}\ ,\nonumber\\
\frac{C_{5-q\ldots 5}}{(\alpha')^{\frac{q+1}{2}}}&=&-{\theta
\hat{\ell}^{1-q}\over h}\Big(\frac{\hat{u}}{\hat{R}}\Big)^{2}\ ,
\label{NS5ymbrane} \\
\frac{B}{\alpha'}&=&\frac{2\hat{R}^{2}\epsilon_{2}}
{\hat{\ell}^{2}} \ , \nonumber\\ \mbox{where}&&
h=1+\theta^{2}\Big(\frac{\hat{u}}{\hat{R}}\Big)^{2}\ ,\quad
\hat{R}^{2}=N \hat{\ell}^2 \ .\nonumber \eea We identify the
\gt{q} coupling, non-commutativity parameter and little string
length scale as

\beq g^2_{\ndgt{q}}={1\over\hat{\ell}^2}\ ,\quad \theta_{{\rm
D}q-\widetilde{\rm GT}} = \theta\hat{\ell}^{q-1}\ ,\quad
\ell_{\rm LST}=\hat{\ell}\ . \la{DqGTtp}\eeq

\subsubsection{Identification of little string lengths}

The following comment on \eq{ODpbrane} and \eq{NS5ymbrane} is in
order: at small $\tilde{r}/\tilde{R}$ or $\hat{u}/\hat{R}$ these
solutions approach the undeformed near horizon geometry of a
NS$5$-brane, which is the supergravity dual of six-dimensional
LST with length scale given by \eq{ODpp} and \eq{DqGTtp}.

These identifications can be understood as follows
\cite{GMSS,harmark2}: independently of whether the NS$5$-brane is
deformed or not, the dilaton always grows large close to an
NS$5$-brane, that is for small $\tilde{r}/\tilde{R}$ or
$\hat{u}/\hat{R}$. In IIB string theory the resulting S-dual
D$5$-brane dynamics is described by six-dimensional Yang-Mills
theory with coupling $g^2_{\rm YM}$, which defines the IIB LST
length scale

\beq \mbox{IIB}\ :\quad \ell_{\rm LST}^2\equiv g_{\rm YM}^2\ .
\la{iiblsl}\eeq From the supergravity dual of the OD$p$ theory
given in \eq{ODpbrane} and the one of the \gt{q} theory given in
\eq{NS5ymbrane} it follows that $g_{\rm YM}^2$ is given by
$\tilde{\ell}^2$ and $\hat{\ell}^2$, respectively, so that
\eq{iiblsl} implies the identifications of the little string
length scales made in \eq{ODpp} and \eq{DqGTtp} for type IIB.

The IIA little strings have a different origin\footnote{We are
thankful for J.P. van der Schaar for pointing this out to us.}.
At low energies the effective $(2,0)$ tensor theory on the
NS$5$-brane flows to a tensor theory on the M-theory five-brane
reduced on a large transverse circle $S^1_T$ with radius $R_T$.
In this picture, the little strings are boundaries of open
membranes that are wrapped around $S^1_T$. The tension of the IIA
little strings is therefore the same as that of the IIA closed
strings, while their strong coupling origin in M-theory is
different. The expression for the tension is given by

\beq \mbox{IIA}\ :\quad \ell^{2}_{\rm LST}\equiv
R_T^{-1}\ell_{11}^3 \ ,\la{iialsl}\eeq where $\ell_{11}$ is the
eleven-dimensional length scale that is fixed in the near horizon
region of the M$5$-brane (the dilaton cancels out as usual). It
follows that $\ell_{\rm LST}$ is given by $\tilde{\ell}$ in
\eq{ODpbrane} and $\hat{\ell}$ in \eq{NS5ymbrane}, which shows
\eq{ODpp} and \eq{DqGTtp} also for type IIA.

In the above we have identified the $\ell_{\rm LST}$ at small
separations/low energies. The LST on the NS$5$-branes decouple,
on the other hand, provided that the local closed string coupling
becomes small (so that IIB little strings are F$1$-strings
`trapped' in the throat of the NS$5$-brane while IIA little
strings are wrapped open M$2$-branes with a degenerated
cylindrical direction). This is expected to happen in the UV
regime, as for the OD$1$ theory, but as we shall see in Section
4, extra care is needed in the case of the OD$3$-LST since the
dilaton approaches a constant value in the UV, leaving the
possibility of having the \od{3} theory on the D$5$-brane.

\section{Equivalence between electric and magnetic\\ deformations}

In the previous section we have obtained near horizon geometries
 through various magnetic and electric near horizon
limits. However, despite the fact that the electric and magnetic
near-horizon limits appear to be quite different, it turns out
that the resulting near horizon regions actually are equivalent,
provided that one relates the fixed parameters. The NCOS dual
(\ref{NCOSbrane}) and the D$q$-GT dual (\ref{Dymbrane}) are
identified as follows ($p=q+2\neq 3$):

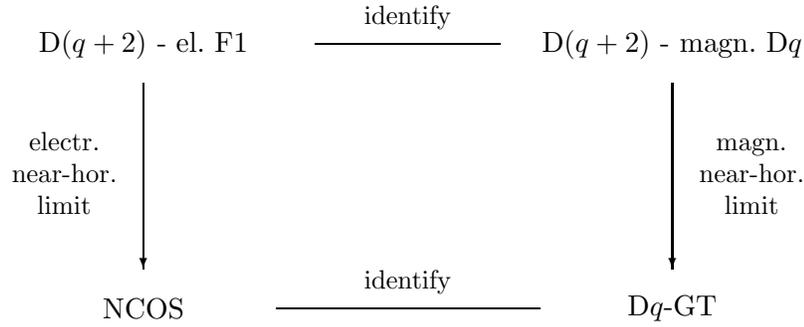
\begin{picture}(200,150)(-90,-20)

\put(0,100){\makebox(0,0){D$(q+2)$ - el. F$1$}}
\put(0,0){\makebox(0,0){NCOS}}
\put(200,100){\makebox(0,0){D$(q+2)$ - magn. D$q$}}
\put(200,0){\makebox(0,0){D$q$-GT}}

\put(0,85){\vector(0,-1){70}} \put(200,85){\vector(0,-1){70}}
\put(65,100){\line(1,0){70}} \put(50,0){\line(1,0){100}}

\put(-30,50){\makebox(0,0){{\small $\begin{array}{c}\mbox{electr.}
\\\mbox{near-hor.}\\\mbox{limit}\end{array}$}}}
\put(230,50){\makebox(0,0){{\small $\begin{array}{c}\mbox{magn.}
\\\mbox{near-hor.}\\\mbox{limit}\end{array}$}}}

\put(100,110){\makebox(0,0){\small identify}}
\put(100,10){\makebox(0,0){\small identify}}

\end{picture}

\begin{equation}\label{id} x^{\mu}=\tilde{x}^{\mu},\quad
u=\frac{\tilde{r}}{\tilde{\ell}^2}\ , \quad
\ell^{p-3}=g\tilde{\ell}^{p-3}\ ,\quad \theta=-\tilde{\ell}^2\ .
\end{equation}
After a change in the sign in the expression for $\theta$, this
transformation also shows the equivalence between the \od{q} dual
(\ref{ODqbrane}) and the NCYM dual (\ref{NCYMbrane}):

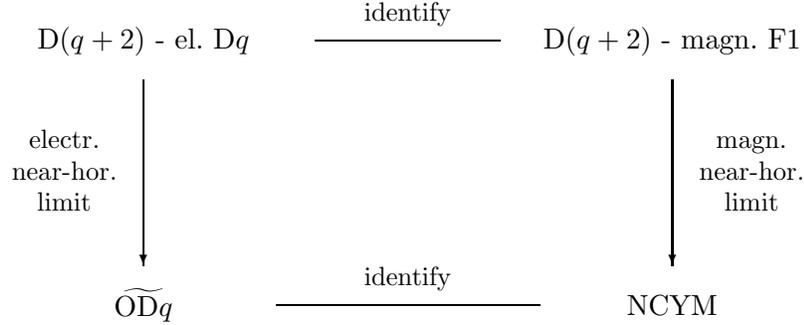
\begin{picture}(200,160)(-90,-30)

\put(0,100){\makebox(0,0){D$(q+2)$ - el. D$q$}}
\put(0,0){\makebox(0,0){\od{q}}}
\put(200,100){\makebox(0,0){D$(q+2)$ - magn. F$1$}}
\put(200,0){\makebox(0,0){NCYM}}

\put(0,85){\vector(0,-1){70}} \put(200,85){\vector(0,-1){70}}
\put(65,100){\line(1,0){70}} \put(50,0){\line(1,0){100}}

\put(100,110){\makebox(0,0){\small identify}}
\put(100,10){\makebox(0,0){\small identify}}

\put(-30,50){\makebox(0,0){{\small $\begin{array}{c}\mbox{electr.}
\\\mbox{near-hor.}\\\mbox{limit}\end{array}$}}}
\put(230,50){\makebox(0,0){{\small $\begin{array}{c}\mbox{magn.}
\\\mbox{near-hor.}\\\mbox{limit}\end{array}$}}}
\end{picture}\\
For $p=q+2=3$ the the identification of $\ell$ and $\tilde{\ell}$
in \eq{id} is replaced by the identification of $g$ in
(\ref{NCOSbrane}) and (\ref{Dymbrane}). In order to identify the
OD$p$ dual (\ref{ODpbrane}) and the \gt{q} dual
(\ref{NS5ymbrane}) we instead do the transformation ($p=4-q$):

\begin{equation}\label{id2} \hat{x}^{\mu}=\tilde{x}^{\mu},\quad
\hat{\ell}=\tilde{\ell}  \ ,\quad
\hat{u}=\frac{\tilde{r}}{\theta}\ ,\quad
\theta=\tilde{g}\tilde{\ell}^{2}\ . \end{equation}

We remark that the brane-coordinates are identified, so the
energies in the related worldvolume theories can be compared
directly without any rescaling. This choice also implies that the
little string theory length scales in the OD$p$ dual
(\ref{ODpbrane}) and the \gt{q} dual (\ref{NS5ymbrane}) are the
same.

Provided that the decoupling of both electric and magnetic
theories makes sense, the above result shows that they are related
in some limit, to which we next turn our attention.

\section{Open D-brane theories on D-branes}

In this section we begin by arguing that the OD$3$ theory is
S-dual to a theory of decoupled, light D$3$-branes on the
D$5$-brane, which we shall denote by \od3, that in turn is dual
to six-dimensional NCYM theory at weak coupling and large
$\theta_{\rm YM}$. T-duality then points to \od{q} theories with
similar properties on the D$(q+2)$-branes for $q=0,1,2,3$.

\subsection{Strongly coupled OD$3$ theory}

The supergravity dual of the six-dimensional OD$3$ theory is given
in (\ref{ODpbrane}) for $p=3$. The OD$3$ theory decouples in the
UV limit $\tilde{r}/\tilde{R}\rightarrow \infty$ and is a theory
of little strings and open D$3$-branes with fixed tensions:

\begin{equation}\label{TLST3}
T_{LST}\equiv {1\over \ell_{LST}^{2}}\ ,\quad T_{\rm OD3}\equiv
\frac{1}{\ell^{4}_{OD3}}=\frac{1}{G^{2}_{\rm OD3}\ell^{4}_{LST}}\ ,
\end{equation}
where $G^{2}_{{\rm OD}3}$ is the OD$3$ coupling constant which is
identified in \eq{ODpp}. For small coupling the OD$3$-theory has a
perturbative description in terms of light little strings in a
special non-commutative geometry giving rise to heavy solitonic
open D$3$-branes. For large $G^{2}_{OD3}$ the situation becomes
reversed, so that the perturbative expansion should be given in
terms of light open D$3$-branes. Indeed, upon S-dualizing the
OD$3$ theory supergravity dual \eq{ODpbrane} we find the \od3
theory ditto \eq{ODqbrane} with the same little string length (we
identify the brane coordinates without any rescaling) provided we
make the identifications

\beq \ell_{\ndod3}=\ell_{{\rm OD}3}\ ,\quad G^2_{\ndod3}={1\over
G^2_{{\rm OD}3}}\ .\la{od3tc}\eeq

In the UV limit, the two configurations given in \eq{ODpbrane} and
\eq{ODqbrane} have identical scaling behavior (with S-dual
parameters), so that it is natural to assume that both the OD$3$
theory and the \od3 theory contain little F$1$ and D$1$-strings,
respectively, as well as open D$3$-branes, with the main
difference that the light objects at weak coupling are the little
strings in OD$3$ theory and the open D$3$-branes in \od3 theory.
Therefore we propose that the S-dual of the OD$3$ theory is the
\od3 theory.

In \cite{GMSS,NS5,NS51} it was argued that OD$3$ is S-dual to
six-dimensional NCYM theory. This is not in contradiction with
the above, but we need to refine the statement as follows: the
NCYM and \od3 theories arise in open F$1$ and D$3$-pictures,
respectively, that are valid in different limits of the parameter
space and provide different perturbative descriptions. From
\eq{id} it follows that the parameters of the two theories are
related as follows:

\beq \theta_{\rm YM}=G_{\ndod3}^{-1}\ell_{\ndod3}^2\ ,\quad g_{\rm
YM}^2=G_{\ndod3}\ell_{\ndod3}^2\ .\la{ncymod3}\eeq The effective
t'Hooft coupling and the effective non-commutativity parameter at
energy scale $E$ are then given by

\bea g_{\rm eff}^2&\equiv& Ng_{\rm
YM}^2E^2=NG_{\ndod3}(E\ell_{\ndod3})^2\ ,\nonumber\\ \quad
\theta_{\rm eff}&\equiv& \theta_{\rm
YM}E^2=G_{\ndod3}^{-1}(E\ell_{\ndod3})^2\ .\eea Thus weakly
coupled \od3 theory arises in the limit of large $\theta_{\rm
YM}$ provided $g^2_{\rm YM}$ becomes small so that $\ell_{\ndod3}$
is kept fixed. At fixed energy scale $E$, this amounts to keeping
$E\ell_{\ndod3}$ fixed while sending $G_{\ndod3}\rightarrow 0$.
Thus \od3 theory and NCYM theory are related at weak (effective)
coupling but in a limit where the perturbative description of the
spatial non-commutativity in the NCYM theory breaks down. The
tension of the NCYM $3$-brane soliton is fixed in the limit and
equal to the tension of the open D$3$-branes:

\beq T_{{\rm NC}3{\rm s}}\equiv {1\over g_{\rm YM}^2\theta_{\rm
YM}}={1\over \ell_{\ndod3}^4}\ .\la{tmp}\eeq Conversely, the NCYM
theory arise in \od3 theory in the limit of small $G^2_{\ndod3}$
and small length scale, i.e. in the weakly coupled low energy
limit, where it is natural to expect an effective field theory
description.

A variant of the above reasoning shows that the \od3 theory may
be a UV-completion of NCYM theory. We then keep $g^2_{\rm eff}$
fixed (instead of sending it to zero), by taking $g_{\rm YM}$ to
zero and the energy scale $E$ to infinity at fixed
$\ell_{\ndod3}$. For large $E\ell_{\ndod3}$ and fixed $g_{\rm
eff}^2$ we see that $G^2_{\ndod3}$ becomes small, which in turn
implies that $\theta_{\rm eff}$ becomes large. Therefore the
perturbative description of the NCYM theory breaks down, as
expected, but since $\ell_{\ndod3}$ remains fixed and
$G_{\ndod3}^2$ small, this points to a well-defined (finite) \od3
theory.

Thus, in summary we propose that OD$3$ and \od3 are S-dual
UV-completions of six-dimensional NCYM theory (with
non-commutativity in two spatial directions).

\subsection{The \od{q} theories ($q=1,2,3$)}

We expect the properties of the \od3 theory and its relation to
NCYM theory to be invariant under T-duality, while the
identification of its strong coupling duals of course changes.
This leads to the notion of \od{q} theories for $q=1,2,3$ as
($q+3$)-dimensional non-gravitational theories containing light
open D$q$-branes that arise in the critical UV limits of the
supergravity duals given by (\ref{ODqbrane}) (the case $q=0$ is
special, and will be discussed separately below). In Section 3 we
showed that the \od{q} supergravity duals \eq{ODqbrane} are
identical to the NCYM duals \eq{NCYMbrane} upon the
identification \eq{id}. It follows that the NCYM parameters
\eq{NCYMp} are related to the \od{q} parameters in \eq{Odqp} as
follows\footnote{Only the relative strengths of the couplings are
important; the precise values of the powers are of course just
artifacts of the identification in \eq{Odqp}. An important issue
for future research is whether it is possible to give an
independent derivation of covariant expressions for the open
D-brane coupling and tension, which are valid in an arbitrary
slowly varying background, that will justify \eq{Odqp}.}:

\beq g^2_{\rm YM}=(G_{\ndod{q}}^2)^{2\over
q+1}\ell_{\ndod{q}}^{q-1}\ ,\quad \theta_{\rm YM}
=(G_{\ndod{q}}^2)^{-2\over q+1} \ell_{\ndod{q}}^2\ .\la{i1}\eeq
The structure of these relations for general $q$ is similar to
that of $q=3$ given in \eq{ncymod3}. The important property is
that $\ell_{\ndod{q}}$ remains fixed and $G^2_{\rm \ndod{q}}$
becomes small in the limit of large $\theta_{\rm YM}$ and small
$g^2_{\rm YM}$. Therefore the relations between NCYM theory and
\od3 theory found in Section 4.1, carry over to the other values
of $q$. In particular, the tension of the NCYM solitons agrees
with the tension of the light open D$q$-branes, i.e. the relation
\eq{tmp} T-dualizes to $\theta_{\rm YM}g^2_{\rm YM}=
\ell^{q+1}_{\ndod{q}}$.

The identification of $G^2_{\ndod q}$ in \eq{Odqp} can be
verified independently in the case of $q=2$ by examining the
uplifting of the \od2 theory supergravity dual, which yields the
OM theory supergravity dual:

\begin{eqnarray}\label{OM}
\frac{ds^{2}_{11}}{\ell_{p}^{2}}&=&\frac{1}{\ell_{\rm OM}^{2}}
\Big[\frac{\tilde{r}^{2}}
{L^{2}}H^{1/3}(-d\tilde{x}^{2}_{0}+d\tilde{x}^{2}_{1}+d\tilde{x}^{2}_{2})
+\frac{L}{\tilde{r}}H^{-2/3}(d\tilde{x}^{2}_{3}+d\tilde{x}^{2}_{4}+
d\tilde{x}^{2}_{5})\nonumber\\ & &+\frac{L}
{\tilde{r}}H^{1/3}(d\tilde{r}^{2}+\tilde{r}^{2}d\Omega^{2}_{4})\Big]\
,\nonumber\\
\frac{A_{012}}{\ell^{3}_{p}}&=&-\frac{1}{\ell^{3}_{\rm OM
}}\frac{\tilde{r}^{3}} {L^{3}} \ , \quad
\frac{A_{345}}{\ell^{3}_{p}}=-\frac{1}{\ell^{3}_{\rm
OM}}\frac{1}{H} \ , \\ H&=&1+\frac{L^{3}}{\tilde{r}^{3}}\ ,\quad
L^{3}= N\ell^{3}_{\rm OM }\ , \nonumber \end{eqnarray} provided
we make the following identification of parameters:

\begin{equation} R_{\rm m}=G^{4\over 3}_{\ndod{2}}\ell_{\ndod2}\
,\quad \ell_{\rm OM }=\ell_{\ndod{2}} \ .\la{od2om} \end{equation}
Hence the radius $R_{\rm m}$ of the magnetic circle on which the
\od 2 theory descends from OM theory indeed becomes small in
units of $\ell_{\ndod2}$ when $G^2_{\ndod2}$ becomes small.

Interestingly, it follows from \eq{od2om} that OM theory can
descend to weakly coupled \od2 theory on a small circle while the
effective OM coupling $E\ell_{\rm OM}$, where $E$ is the
six-dimensional energy scale, remains finite. This suggests that
the OM theory interactions are reminiscent of those in the weakly
coupled \od2 theory.

The strong coupling limit of the \od{1} theory will be discussed in Section 5.

\subsection{The \od0 theory}

The three-dimensional \od0 theory arises in the IR limit of the
supergravity dual \eq{ODqbrane} and is dual to NCYM theory and
the $SO(8)$-invariant conformal theory on the M-theory membrane.
The lift of the \od0 supergravity dual \eq{ODqbrane} to M-theory
gives the near horizon region of a M$2$-MW bound state (the
membrane is transverse to the wave but smeared in the $11$
direction):

\begin{eqnarray}\label{M2w}
\frac{ds^{2}}{\ell_{p}^{2}}&=&\frac{1}{\ell_{{\rm Pl}
}^{2}}H^{1/3}\Big (\frac{\tilde{r}}
{\tilde{R}}\Big)^{5}\Big[-d\tilde{x}_{0}^{2}+H^{-1}\Big(\frac{\tilde{R}}
{\tilde{r}}\Big)^{5}(d\tilde{x}_{1}^{2}+d\tilde{x}_{2}^{2}) \nonumber\\
&+&\Big(\frac{\tilde{R}}{\tilde{r}}
\Big)^{5}(d\tilde{r}^{2}+\tilde{r}^{2}d\Omega^{2}_{6})+
H^{-1}\Big(\frac{\tilde{R}}{\tilde{r}}
\Big)^{10}\Big(d\tilde{x}^{11}+\Big(\frac{\tilde{r}}
{\tilde{R}}\Big)^{5}d\tilde{x}^{0}\Big)^{2}\Big]\ , \nonumber\\
\frac{A_{012}}{\ell_{p}^{3}}&=&\frac{1}{\ell_{{\rm Pl
}}^{3}}\frac{1}{H}\ ,\quad
\frac{A_{12,11}}{\ell_{p}^{3}}=-\frac{1}{\ell_{{\rm
Pl}}^{3}}\frac{1}{H}\ ,
\nonumber\\
H&=&1+\Big(\frac{\tilde{R}}{\tilde{r}}\Big)^{5}\ , \quad
\tilde{R}^{5}= \frac{N\ell_{{\rm Pl}}^{6}}{R_{11}}\ ,
\end{eqnarray}
with eleven dimensional length scale $\ell_{\rm Pl}$ and radius
$R_{11}$ given by:

\begin{equation}\label{M2W} \ell_{\rm Pl}=g^{1\over
3}\tilde{\ell}=G^{-{4\over 3}}_{\ndod{0}}\ell_{\ndod{0}} \ ,\quad
R_{11}=g\tilde{\ell}=\ell_{\ndod0}\ .\end{equation}

For large $\tilde{r}/\tilde{R}$ the eleven-dimensional geometry
is dominated by a wave propagating in the light-like $11$
direction in a flat background with Planck length $\ell_{\rm
Pl}$. The ten-dimensional UV region is dominated by the
D$0$-brane charge. In the gauge $X^0=\tau$ the D$0$-brane action
reduces in the UV limit to the quantum mechanical kinetic term
${1\over 2}M_{\ndod0}\dot{X}^2$ where $\dot{X}$ is the
nine-dimensional velocity and the D$0$-mass is given by

\beq M_{\ndod0}=(\ell_{\ndod0})^{-1}\ .\la{d0m}\eeq
This is the supergravity dual description \cite{BGLHKS} (the
BGLHKS limit) of the original DLCQ limit of M-theory
\cite{DKPS,SSS} (the DKPS and SSS limits)\footnote{For a review
see e.g. \cite{polch}.}.  Thus the UV limit of the \od0 theory
with N units of D$0$-brane charge is a DLCQ compactification of
M-theory with $N$ units of DLCQ momentum on a light-like circle
with radius $R_{11}$, in the presence of transverse M$2$-branes.
This is similar to the strong coupling limit of the OD$0$ theory
\cite{GMSS} (change the M$2$-brane to an M$5$-brane). However,
unlike the usual undeformed matrix model which have only a
dimensionful coupling, the \od0 matrix model inherits the
dimensionless coupling $G^2_{\ndod0}$, which is encoded in the
transverse two-brane. From \eq{M2W} it follows that $\ell_{\rm
Pl}$ becomes large when $G^2_{\ndod0}<<1$ and $\ell_{\ndod0}$ is
fixed, while it becomes small, leading to eleven-dimensional
supergravity, at strong \od0 coupling, $G^2_{\ndod0}>>1$, and
fixed $\ell_{\ndod0}$. In the limit of large $N$, $G^2_{\ndod0}$
and $\ell_{\ndod0}$ (at fixed energy) we recover the BFSS limit,
i.e. $R_{11}\rightarrow \infty$ at fixed $\ell_{\rm Pl}$ and
light-like momentum $N/R_{11}$, which leads to weakly coupled
supergravity with finite but small derivative corrections.

For small $\tilde{r}/\tilde{R}$, the geometry is dominated by the
two-branes, so that the open D$0$-brane physics becomes
three-dimensional, instead of strongly coupled quantum mechanics.
In the open string picture this gives rise to three-dimensional
NCYM theory with fixed $g^2_{\rm YM}$ and $\theta_{\rm YM}$ given
by \eq{i1} (the non-commutativity is independent of
$\tilde{r}/\tilde{R}$ and non-zero in the IR even though the NS
potential vanishes in there). Its strong coupling dual is the
$SO(8)$-invariant conformal theory (the $N=8$ singleton) on the
M$2$-brane \cite{GMSS}. This theory has a critical mass-term
determined by the length scale $\ell_{\rm PL}$ of the
supergravity dual \eq{M2w}. Dualizing one of its scalars, by
first identifying it with a circle with radius $R_{11}$, leads to
a NCYM theory with supergravity dual \eq{NCYMbrane} provided that:

\beq \ell_{SO(8)}\equiv\ell_{\rm Pl}=g^{2\over 3}_{\rm
YM}\theta_{\rm YM}^{2\over 3}\ ,\quad R_{11}=g^2_{\rm
YM}\theta_{\rm YM}\ .\la{so8ym}\eeq Thus at strong coupling,
$g_{\rm YM}^2>>E$ where $E$ is a fixed energy, and small
$\theta_{\rm YM}$, such that $g_{\rm YM}\theta_{\rm YM}$ is fixed,
the NCYM theory goes over into the $SO(8)$ theory with large
$R_{11}$ and fixed $\ell_{SO(8)}$, that is fixed effective
coupling $E\ell_{SO(8)}$.

In the limit of large $\theta_{\rm YM}$, small $g^2_{\rm YM}$ and
fixed NCYM soliton mass $(g^2_{\rm YM}\theta_{\rm YM})^{-1}$ both
NCYM and $SO(8)$ theories break down, and the perturbative
description is given by the \od0 theory which has an expansion in
$G^2_{\ndod0}$ at fixed $\ell_{\ndod0}$. The NCYM solitons remain
weakly coupled, provided the strong NCYM interactions at long
wave-length are under control, and their mass is equal to the
D$0$-mass given in \eq{d0m}, which is constant under the flow
from UV down to the IR. We propose that the \od0 theory is based
on critical open D$0$-branes, which are identified as D$0$-branes
that are created on the D$2$-brane and remains bound to it (in a
weak, sub-critical electric field). The strong coupling dual of
the \od0 theory is the $SO(8)$ theory. For small $G^2_{\ndod0}$
and fixed $\ell_{\ndod0}$ the $SO(8)$ theory becomes strongly
coupled. On the other hand, for large $G^2_{\ndod0}$ and fixed
$\ell_{\ndod0}$ the $SO(8)$ length scale $\ell_{SO(8)}$ becomes
small and $R_{11}$ fixed, while if $\ell_{\ndod0}$ is taken to be
large (in units of a fixed energy) then $\ell_{SO(8)}$ is fixed
and $R_{11}$ becomes large.

\section{Non-commutative generalized gauge and tensor \\ theories on D-branes and
NS$5$-branes}

In the previous section we examined a new type of RR-electric
theories on D-branes, the \od{q} theories, and found that they
are related to NCYM theories at weak coupling and large
$\theta_{\rm YM}$ and to other theories at large coupling and
fixed length scale. In this section we show that the arrangement
of the NCOS, NCYM, OD$q$ and \od{q} theories into duality webs
requires another set of decoupled theories -- the D$q$-GT and the
\gt{q} -- that are perturbative in the corners of the parameter
space corresponding to the supergravity duals \eq{Dymbrane} and
\eq{NS5ymbrane} with fixed parameters \eq{DqGTp} and \eq{DqGTtp},
respectively. These supergravity solutions are obtained in
magnetic near horizon limits of D-branes and NS$5$-branes with
fixed magnetic RR fluxes in the UV, and as was shown in Section 3
they are equivalent to the NCOS and OD$p$ duals \eq{NCOSbrane}
and \eq{ODpbrane}. This means that the new theories have fixed
spatial non-commutativity in limits where the temporal
non-commutativity length scales of the NCOS and OD$p$ theories
become large.

\subsection{Generalized gauge theories on D-branes
 (D$q$-GT)}

The D$q$-GT is defined by the supergravity dual \eq{Dymbrane} and
has coupling $g^{2}_{Dq-GT}$ and spatial non-commutativity
parameter $\theta_{{\rm D}q{\rm -GT}}$, which measures the
strength of an anti-symmetric tensor of rank $(q+1)$ and of
dimension length to the power $q+1$, given in \eq{DqGTp}. From
\eq{NCOSp} and \eq{id} it follows that:

\begin{equation}\label{OSGT}
g^2_{Dq-GT}=G_{\rm OS}^2(\alpha'_{\rm eff})^{q-1\over2}\ ,\quad
\theta_{Dq-GT}=G^{2}_{OS}(\alpha'_{\rm eff})^{\frac{q+1}{2}}\ .
\end{equation}
This relation together with the assumption that the NCOS theory
should be related to D$q$-GT at weak coupling implies that the
D$q$-GT (if it exists) must arise in the $\alpha'_{\rm
eff}\rightarrow \infty$, or high energy, limit of NCOS at weak
coupling $G^2_{\rm OS}\rightarrow 0$ such that $\theta_{Dq-GT}$
is kept fixed. In this limit the SYM sector of the NCOS theory,
whose Yang-Mills coupling is given by

\beq g^2_{\rm YM}\equiv G^2_{\rm OS}(\alpha'_{\rm eff})^{q-1\over
2}=g^2_{Dq-GT}\ ,\la{NCOSym}\eeq  therefore remains interacting,
which motivates the term `generalized gauge theory'.

The perturbative formulation of ordinary (commutative) open
string theory is not known in this limit, and one may argue that
it simply consists of some kind of degenerate open string world
sheets. However, we believe that the NCOS theory may stand a
better chance, because the supergravity dual description
indicates that the theory could be described in terms of some
structure which encodes the fixed non-commutativity parameter
$\theta_{{\rm D}q-{\rm GT}}$. For $q=2,3$ this type of
non-commutativity is naturally represented on $(q-1)$-loop space;
see for example \cite{us1} for a similar discussion in the case
of the M-theory membrane. This is indicative of a non-commutative
generalization of the ordinary loop space covariant derivatives.
A preliminary analysis hints that such a structure indeed
accommodates a hierarchy of massless gauge fields in $q+3$
dimensions, as one might anticipate in the above limit. If we let
$G_{\rm loop}$ be the gauge coupling and assume that at very low
energies the resulting theory contains a SYM sector, then a
simple dimensional analysis shows that $g^2_{\rm YM}=G^2_{\rm
loop}(\theta_{{\rm D}q-{\rm GT}})^{q-1\over q+1}$. From
\eq{NCOSym} it then follows that $G^{2}_{\rm
loop}=(G^{2}_{OS})^{\frac{2}{q+1}}$ so that $G_{\rm loop}$ is
indeed small in the limit when $G^2_{\rm OS}$ is small.

For $q=2,3$ the D$q$-GT coupling is dimensionful and we shall
assume that it sets the energy scale of the D$q$-GT. From
\eq{OSGT} it follows that if the D$q$-GT non-commutativity scale,
set by the parameter $\theta_{Dq-GT}$, is kept fixed in the limit
of large $\alpha'_{\rm eff}$ and small $G^2_{\rm OS}$ at fixed
energy $E$, then the effective D$q$-GT coupling
$g^2_{Dq-GT}E^{q-1}$ becomes small. In particular, this means
that the effects of the non-commutativity become visible at
energies where the effective D$q$-GT coupling still is small.
Thus the D$q$-GT is expected to have a formulation at long
wave-lengths in terms of some weakly coupled field theory
construction, such as the loop-space gauge theory discussed above.

The case $q=0$ appears to be special. It is not clear to us how to
interpret the vectorial non-commutativity parameter in terms of a
non-commutative structure on some space. We shall therefore
simply define the D$0$-GT as the large $\alpha'_{\rm eff}$ and
small $G^2_{\rm OS}$ limit of three-dimensional NCOS theory.

\subsection{Extended duality webs}

We next turn to the issue of identifying the strong coupling
limits of the D$q$-GT and arrange it together with the other
non-commutative theories into duality diagrams involving limits
of strong coupling as well as large non-commutativity length
scale.

\subsubsection{Four-dimensional
non-commutative theories}

We have four different four-dimensional non-commutative theories
on the D$3$-brane: the NCYM and NCOS theories based on weakly
coupled open F$1$-string interactions at fixed non-commutativity
length scale; the \od1 theory which has a $G^{2}_{\ndod
1}=g^2_{\rm YM}$ expansion at fixed $\ell^2_{\ndod1}=\theta_{\rm
YM}g^{2}_{\rm YM}$; and the D$1$-GT which has a $G^2_{\rm
OS}$-expansion at fixed $\theta_{D1-GT}=\alpha'_{\rm eff}G^2_{\rm
OS}$. Upon combining the relations \eq{i1} and \eq{OSGT} with the
usual S-duality relations between NCOS and NCYM theory, which read

\beq g_{\rm YM}^2={1\over G_{\rm OS}^2}\ ,\quad \theta_{\rm YM}
=G_{\rm OS}^2\alpha'_{\rm eff}\ ,\la{NCYMNCOS}\eeq we find that
at strong coupling and fixed length scales, the NCOS theory go
over into \od1 theory, as follows from

\begin{equation}\label{SD1OS} G^{2}_{\rm
OS}=\frac{1}{G^{2}_{\ndod{1}}}\ ,\quad \alpha'_{\rm eff}=
\ell^{2}_{\ndod{1}}\ , \end{equation} and NCYM theory into
D$1$-GT, as follows from:

\begin{equation} g^{2}_{YM}=\frac{1}{g^{2}_{D1-GT}}\ , \quad
\theta_{\rm YM} =\theta_{D1-GT}\ . \la{ymd1gt}\end{equation} In
going from NCYM theory to D$1$-GT theory the non-commutativity
parameter is held fixed so that the NCYM solitons become
massless, which is equivalent with the D$1$-GT being the
tensionless limit of the NCOS theory. Of course, NCOS go over
into NCYM theory in the limit of strong coupling and small
$\alpha'_{\rm eff}$ such that $\theta_{\rm YM}$ is fixed
\cite{gopa}. This can be summarized in the following duality
diagram:

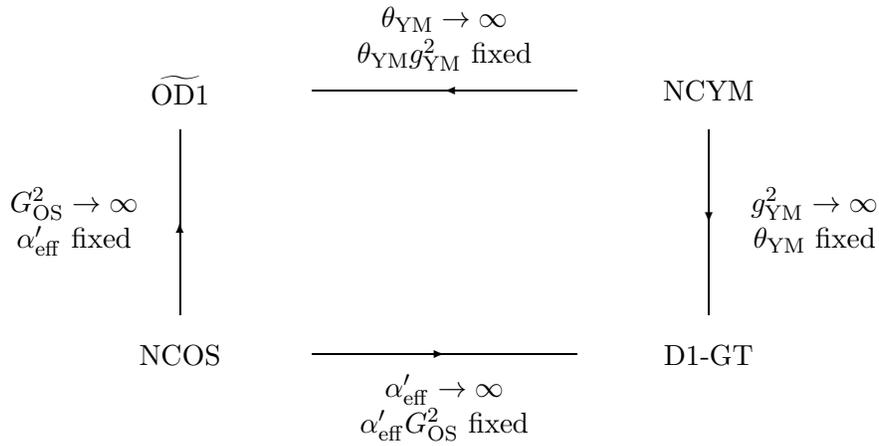
\begin{picture}(200,190)(-100,-50)

\put(0,100){\makebox(0,0){\od1}} \put(0,0){\makebox(0,0){NCOS}}
\put(200,100){\makebox(0,0){NCYM}}
\put(200,0){\makebox(0,0){D$1$-GT}}

\put(0,15){\line(0,1){70}\vector(0,1){35}}
\put(200,85){\line(0,-1){70}\vector(0,-1){35}}
\put(50,100){\line(1,0){100}}\put(150,100){\vector(-1,0){50}}
\put(50,0){\line(1,0){100}}\put(50,0){\vector(1,0){50}}

\put(-40,50){\makebox(0,0){$\begin{array}{c}G^2_{\rm
OS}\rightarrow\infty\\\alpha'_{\rm eff}\
\mbox{fixed}\end{array}$}}
\put(240,50){\makebox(0,0){$\begin{array}{c}g_{\rm
YM}^2\rightarrow\infty\\\theta_{\rm YM}\
\mbox{fixed}\end{array}$}}
\put(100,-20){\makebox(0,0){$\begin{array}{c}\alpha'_{\rm
eff}\rightarrow\infty\\\alpha'_{\rm eff}G^2_{\rm OS}\
\mbox{fixed}\end{array}$}}
\put(100,120){\makebox(0,0){$\begin{array}{c}\theta_{\rm
YM}\rightarrow \infty\\\theta_{\rm YM}g^2_{\rm YM}\
\mbox{fixed}\end{array}$}}

\end{picture}

There is an important asymmetry in the diagram, so that both
types of strings are heavy in NCYM theory, whereas one type is
light and the other one heavy in D$1$-GT. Moreover, as already
mentioned in the Introduction, the IIB $SL(2,{\sf Z})$ does not
map the weakly coupled supergravity duals of the lower part of
the diagram into the the weakly coupled supergravity duals of the
upper part of the diagram, so that it would indeed be artificial
to assume that the diagram would be self-dual in the sense that
the lower half could be identified with the upper half (of
course, when the coupling becomes large in the lower part of the
diagram, then the IIB $SL(2,{\sf Z})$ maps it to the upper part
and vice versa).

\subsubsection{Five-dimensional non-commutative theories}

In five dimensions we can organize the non-commutative theories as
follows:

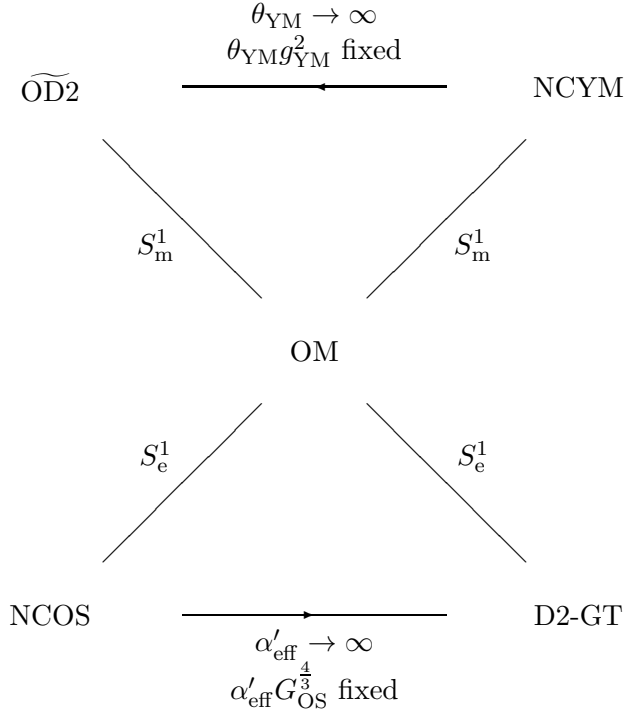
\begin{picture}(200,290)(-100,-50)

\put(0,200){\makebox(0,0){\od2}} \put(0,0){\makebox(0,0){NCOS}}
\put(200,200){\makebox(0,0){NCYM}}
\put(200,0){\makebox(0,0){D$2$-GT}}
\put(100,100){\makebox(0,0){OM}}

\put(20,20){\line(1,1){60}} \put(180,20){\line(-1,1){60}}
\put(20,180){\line(1,-1){60}} \put(180,180){\line(-1,-1){60}}

\put(50,200){\line(1,0){100}}\put(150,200){\vector(-1,0){50}}
\put(50,0){\line(1,0){100}}\put(50,0){\vector(1,0){50}}

\put(100,-20){\makebox(0,0){$\begin{array}{c}\alpha'_{\rm
eff}\rightarrow\infty\\\alpha'_{\rm eff} G^{4\over 3}_{\rm OS}\
\mbox{fixed}\end{array}$}}
\put(100,220){\makebox(0,0){$\begin{array}{c}\theta_{\rm
YM}\rightarrow \infty\\\theta_{\rm YM}g^2_{\rm YM}\
\mbox{fixed}\end{array}$}} \put(40,60){\makebox(0,0){$S^1_{\rm
e}$}} \put(160,60){\makebox(0,0){$S^1_{\rm e}$}}
\put(40,140){\makebox(0,0){$S^1_{\rm m}$}}
\put(160,140){\makebox(0,0){$S^1_{\rm m}$}}

\end{picture}\\
Here $S^1_{\rm m}$ and $S^1_{\rm e}$ denote electric and magnetic
circles in OM theory. The relations between the parameters of the
electric reduction of OM theory and the ones of NCOS theory and
D$2$-GT are given by

\beq \ell_{\rm OM}=G^{2/3}_{OS}\sqrt{\alpha'_{\rm eff}}\ ,\quad
R_{\rm e}=G^{2}_{OS} \sqrt{\alpha'_{\rm eff}}\ ,\eeq

\begin{equation}
\ell_{\rm OM}=\theta_{D2-GT}^{1/3}\ , \quad R_{e}=g^{2}_{D2-GT}
\ .\end{equation} The relations in the
magnetic case follow from \eq{i1} and \eq{od2om}, which we
collect here for completeness:

 \beq \ell_{\rm OM}=g^{2\over 3}_{\rm YM}\theta^{1\over 3}_{\rm YM}\
,\quad R_{\rm m}=g^2_{\rm YM}\ ,\eeq

\beq \ell_{\rm OM}=\ell_{\ndod2}\ ,\quad R_{\rm m}=G^{4\over
3}_{\ndod2}\ell_{\ndod2}\ .\eeq Thus the reductions of OM theory
at fixed energy $E$ to D$2$-GT and \od2 keeps the six-dimensional
effective coupling $g_{\rm OM,eff}^2\equiv E\ell_{\rm OM}$ fixed,
as opposed to the reductions to the NCOS and NCYM theories where
$g_{\rm OM,eff}^2\rightarrow 0$ at weak string coupling.

\subsubsection{Three-dimensional non-commutative theories}

In \cite{GMSS} it was conjectured that three-dimensional NCOS
theory lifts to the M$2$-brane $SO(8)$ theory in the limit of
strong coupling. The D$0$-GT theory has the same supergravity
dual as the NCOS theory and obey the relation (\ref{OSGT}) with
$q=0$. We therefore conjecture that the D$0$-GT theory has the
same strong coupling limit as three-dimensional NCOS theory. The
$SO(8)$ theory also governs the strong coupling behavior of
three-dimensional NCYM theory. In Section 4.3 we proposed to
extend this to the \od0 theory. Thus we organize the
non-commutative theories in three dimensions as follows:

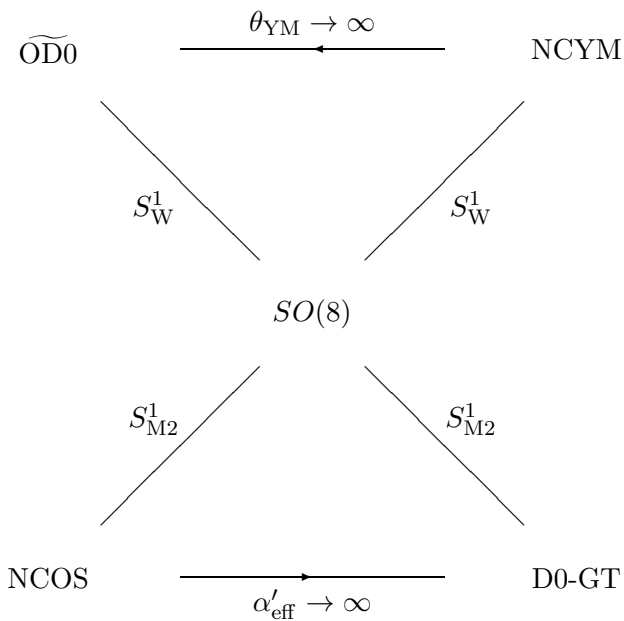
\begin{picture}(200,260)(-100,-30)

\put(0,200){\makebox(0,0){\od0}} \put(0,0){\makebox(0,0){NCOS}}
\put(200,200){\makebox(0,0){NCYM}}
\put(200,0){\makebox(0,0){D$0$-GT}}
\put(100,100){\makebox(0,0){$SO(8)$}}

\put(20,20){\line(1,1){60}} \put(180,20){\line(-1,1){60}}
\put(20,180){\line(1,-1){60}} \put(180,180){\line(-1,-1){60}}

\put(50,200){\line(1,0){100}}\put(150,200){\vector(-1,0){50}}
\put(50,0){\line(1,0){100}}\put(50,0){\vector(1,0){50}}

\put(100,-10){\makebox(0,0){$\alpha'_{\rm eff}\rightarrow\infty$}}
\put(100,210){\makebox(0,0){$\theta_{\rm YM}\rightarrow \infty$}}
\put(40,60){\makebox(0,0){$S^1_{{\rm M}2}$}}
\put(160,60){\makebox(0,0){$S^1_{{\rm M}2}$}}
\put(40,140){\makebox(0,0){$S^1_{\rm W}$}}
\put(160,140){\makebox(0,0){$S^1_{\rm W}$}}
\end{picture}\\
Here $S^{1}_{W}$ amounts to a boost of the M$2$-brane followed by
a transverse reduction in the resulting wave direction.
$S^{1}_{{\rm M}2}$ denotes a reduction along a (skew) direction
in a smeared M$2$-M$2$ bound state. The relations between the
$SO(8)$ theory parameters and the non-commutative theories in
type IIA are given by \eq{M2W} and \eq{so8ym}. As in the
five-dimensional case we find that the reductions to the
D$0$-brane theories keep the effective $SO(8)$ coupling
$E\ell_{SO(8)}$ fixed at fixed energy $E$.

\subsection{Generalized gauge theories on the type IIB NS$5$-brane}

On the IIB five-branes the situation is not yet quite symmetric;
after the inclusion of the \od3 theory and the D$3$-GT we have
two `triplets': i) 1) NCOS theory (NS-electric D$5$ in the open
F$1$-picture); 2) D$3$-GT (RR-magnetic D$5$-brane in the open
D$3$-brane picture); and 3) OD$1$ theory (RR-electric IIB
NS$5$-brane in LST-picture), and ii)  1') NCYM theory
(NS-magnetic D$5$-brane in the open F$1$-picture); 2') \od3
theory (RR-electric D$5$-brane in the open D$3$-brane picture);
and 3') OD$3$ theory (RR-electric IIB NS$5$-brane in the
LST-picture).

The triplet (i) is completed by a theory, which we shall denote by
\gt3, that arises in the open D$3$-brane picture on a RR-magnetic
NS$5$-brane. We shall assume that \gt3 is some limit of OD$1$ at
weak coupling $G^2_{{\rm OD}1}<<1$. From \eq{DqGTtp} and \eq{id2}
it follows that the \gt3 theory has coupling (with dimension
energy to the power two) and spatial non-commutativity parameter
(which is an anti-symmetric tensor of rank $4$ and dimension
length to the power $4$) given by

\beq g^2_{\ndgt3}={1\over  \ell_{\rm LST}^2}\ ,\quad
\theta_{\ndgt3}=G^2_{{\rm OD}1}\ell_{\rm LST}^4\ .\la{lstgt3}\eeq
Hence the \gt3 arises in the $\ell_{\rm LST}\rightarrow \infty$
limit of OD$1$, at weak OD$1$ coupling and fixed
$\theta_{\ndgt3}$. Importantly, the OD$1$ solitons become heavy
in this limit, since their length scale $G_{{\rm OD}1}\ell_{\rm
LST}$ becomes small. The positive dimension of the coupling means
that the \gt3 becomes strongly coupled at energies below
$g_{\ndgt3}=\ell_{\rm LST}^{-1}$, which is due to little string
effects. Thus, for energies above $\ell^{-1}_{\rm LST}$ we expect
to describe \gt3 in some language free of little strings. Hence,
for large $\ell_{\rm LST}$, small $G^2_{{\rm OD}1}$ and fixed
$\theta_{\ndgt3}$ there is an interval of energies
$g_{\ndgt3}<<E<<\theta_{\rm \ndgt3}^{-{1\over 4}}$ where the
little string effects are unimportant and the effective
non-commutativity $E^4\theta_{\ndgt3}$ is small, so that it can
be described perturbatively using some generalized gauge theory
based on open D$3$-branes. Thus \gt3 differs from OD$1$ theory,
whose temporal non-commutativity becomes visible first at strong
effective coupling at energies above the little string scale.

We therefore expect the structure of the \gt3 to resemble that of
the D$3$-GT. Indeed, their supergravity duals, \eq{Dymbrane} and
\eq{NS5ymbrane}, are related by S-duality without any change in
the magnetic RR 4-form potential, which implies

 \beq
\theta_{\ndgt3}=\theta_{D3-GT}\ ,\quad g^2_{\ndgt3}={1\over
g^2_{D3-GT}}\ .\la{d3gtdual}\eeq In order to interpret this, we
recall that the generalized gauge theory description of D$3$-GT
is perturbative when
$E<<\theta_{D3-GT}^{-{1\over4}}<<g^{-1}_{D3-GT}$. From
\eq{d3gtdual} it follows that when the generalized gauge theory
description of D$3$-GT breaks down at
$\theta_{D3-GT}^{-{1\over4}}> g^{-1}_{D3-GT}$ then the
corresponding formulation of \gt3 becomes well-defined. The
generalized gauge theory descriptions of D$3$-GT and \gt3 are
therefore S-dual at fixed energy and non-commutativity parameter.
We summarize as follows:

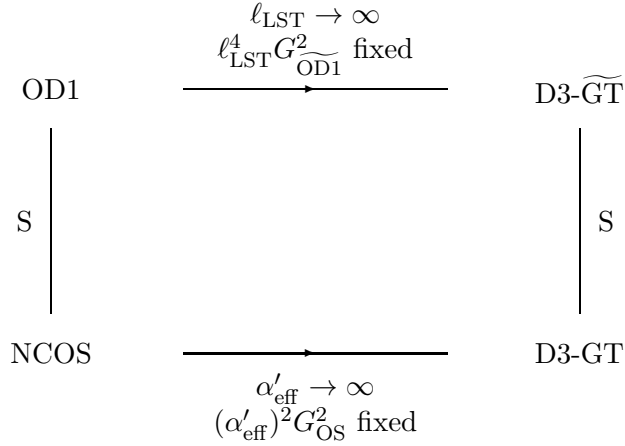
\begin{picture}(200,200)(-100,-50)

\put(0,100){\makebox(0,0){OD$1$}} \put(0,0){\makebox(0,0){NCOS}}
\put(200,100){\makebox(0,0){\gt3}}
\put(200,0){\makebox(0,0){D$3$-GT}}

\put(0,15){\line(0,1){70}} \put(200,85){\line(0,-1){70}}
\put(50,100){\line(1,0){100}}\put(50,100){\vector(1,0){50}}
\put(50,0){\line(1,0){100}}\put(50,0){\vector(1,0){50}}

\put(-10,50){\makebox(0,0){S}} \put(210,50){\makebox(0,0){S}}
\put(100,-20){\makebox(0,0){$\begin{array}{c}\alpha'_{\rm
eff}\rightarrow\infty\\(\alpha'_{\rm eff})^2 G^2_{\rm OS}\
\mbox{fixed}
\end{array}$}}
\put(100,120){\makebox(0,0){$\begin{array}{c}\ell_{\rm
LST}\rightarrow \infty\\\ell^4_{\rm LST}G^2_{\ndod1}\
\mbox{fixed}\end{array}$}}

\end{picture}

Similarly, the triplet (ii) is completed by \gt1 that is S-dual
to NCYM theory at fixed length and the tensionless little string
limit of OD$3$ at weak OD$3$-coupling:

\beq \theta_{\ndgt1}=\theta_{\rm YM}=G_{{\rm OD}3}^2\ell_{\rm
LST}^2\ ,\quad g^2_{\ndgt1}={1\over g^2_{\rm YM}}
={1\over\ell_{\rm LST}^2}\ .\la{d1gtt}\eeq The reason \gt1 is
S-dual to NCYM theory, and not another generalized gauge theory
as was the case for \gt3, is related to that the solitonic
D$3$-branes become massless in the tensionless little string
limit of OD$3$. This is equivalent to the vanishing of the
tension \eq{tmp} of the non-commutative solitons in the
six-dimensional NCYM theory in the limit of large $g^2_{\rm YM}$
and fixed $\theta_{\rm YM}$. Thus, from \eq{d1gtt} it follows
that we can identify \gt1 with six-dimensional NCYM theory in the
limit of large $g^2_{\rm YM}$ at fixed $\theta_{\rm YM}$. We
summarize as follows:

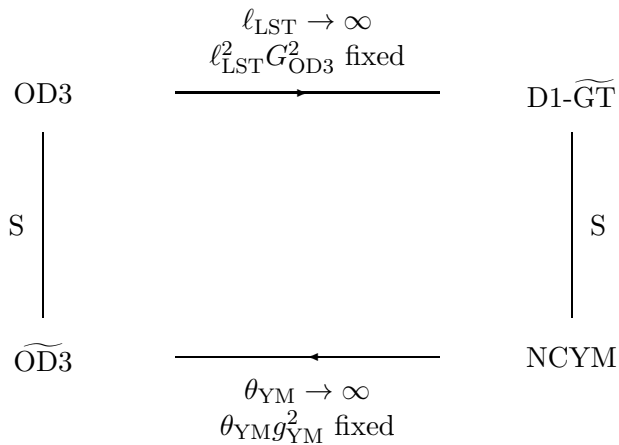
\begin{picture}(200,190)(-100,-50)

\put(0,100){\makebox(0,0){OD$3$}} \put(0,0){\makebox(0,0){\od3}}
\put(200,100){\makebox(0,0){\gt1}}
\put(200,0){\makebox(0,0){NCYM}}

\put(0,15){\line(0,1){70}} \put(200,15){\line(0,1){70}}
\put(50,100){\line(1,0){100}}\put(50,100){\vector(1,0){50}}
\put(50,0){\line(1,0){100}}\put(150,0){\vector(-1,0){50}}

\put(-10,50){\makebox(0,0){S}} \put(210,50){\makebox(0,0){S}}
\put(100,-20){\makebox(0,0){$\begin{array}{c}\theta_{\rm
YM}\rightarrow\infty\\\theta_{\rm YM}g^2_{\rm YM}\
\mbox{fixed}\end{array}$}}
\put(100,120){\makebox(0,0){$\begin{array}{c}\ell_{\rm
LST}\rightarrow \infty\\\ell^2_{\rm LST}G^2_{{\rm OD}3}\
\mbox{fixed}\end{array}$}}

\end{picture}

\subsection{Generalized tensor theories on the type IIA NS$5$-brane}

T-dualizing the D$q$-GT ($q=1,3$) on the IIB NS$5$-brane leads to
the \gt{q} ($q=0,2,4$) on the IIA NS$5$-brane based on the
supergravity duals \eq{NS5ymbrane}. These theories have spatial
non-commutativity and coupling with positive energy dimension
given by \eq{DqGTp}. For energies larger than the coupling and
smaller than the non-commutative length scale we expect these
theories to have perturbative formulations in terms of some
tensionless extensions of the $(2,0)$ tensor theory incorporating
loop space non-commutativity.

The supergravity duals of \gt{q} and OD$(4-q)$ theory are
related, as shown in Section 3, leading to relations between their
parameters analogous to \eq{lstgt3}. In the limit of large
$\ell_{\rm LST}$ and fixed $\theta_{\ndgt{q}}$ the solitonic
D$(4-q)$-branes of the OD$(4-q)$ theory become tensionless when
$q=0$ and heavy when $q=4$, while the solitonic D$2$-branes of
the OD$2$ theory remain in the spectrum with a finite tension
given by the non-commutative length scale:

\beq T_{{\rm OD}2}={1\over\tilde{g}\tilde{\ell}^3}={1\over
\theta_{\ndgt2}}\ ,\eeq where we have used \eq{ODpp}, \eq{DqGTtp}
and \eq{id2}. The generalized tensor theory becomes strongly
coupled when $g_{\ndgt{q}}\theta_{\ndgt{q}}^{1\over q+1}>1$,
since then there is no energy interval where both little string
and non-commutative effects are weak.

The supergravity dual of \gt0 and the OD$4$ theory is a reduction
of the near horizon region of a (smeared) M5-M5 bound state along
a skew direction. We therefore expect these theories to be
governed at strong coupling by the M$5$-brane $(2,0)$ tensor
multiplet. The supergravity duals of \gt4 and the OD$0$ theory
descend from a boosted M$5$-brane with a transverse wave. The
wave provides the electric D$0$-brane charge, which becomes heavy
in the limit of small radius.

The case of $q=2$ requires extra care. The six-dimensional
worldvolume has self-dual RR three-form flux. The electric near
horizon limit yields the OD$2$ supergravity dual \eq{ODpbrane}
and the magnetic near horizon limit yields the \gt2 supergravity
dual \eq{NS5ymbrane}. These are equivalent\footnote{Thus, if we
were to use an open D$2$-brane description, their fluctuations
around the NS$5$-brane would form a `mixed state' of fluctuations
in both electric and magnetic directions, similar to the
$(p,q)$-string fluctuations in the IIB theory. The perturbative
description of the OD$2$ theory and the \gt{2} would then have to
be extracted from a single sector treated as fundamental, say in
a semi-classical expansion around a classical solution.}, as we
have shown in Section 3. Thus the OD$2$ and \gt2 theories are
perturbative descriptions valid for small and large $\ell_{\rm
LST}$, respectively, at fixed energy (for large $\ell_{\rm LST}$
we take small dimensionless coupling so that $\theta_{\ndgt{2}}$
is fixed). For energies $g_{\ndgt2}<<E<<\theta_{{\rm
OD}2}^{-{1\over3}}=\ell_{{\rm OD}2}^{-1}$ the \gt2 theory is
expected to have a perturbative expansion as a generalized tensor
theory with weak spatial non-commutativity. When
$g_{\ndgt2}>\theta_{{\rm OD}2}^{-{1\over3}}$ the theory becomes
strongly coupled. The uplift of the OD$2$ and \gt2 supergravity
duals is identified with the OM theory supergravity dual reduced
on a transverse circle $S^1_T$ \cite{harmark2}:

\bea \ell_{\rm OM}&=&G^{2\over 3}_{{\rm OD}2}\ell_{\rm LST}=
\theta^{1\over 3}_{\ndgt2}\ ,\nonumber\\ R_T&=&G^2_{{\rm
OD}2}\ell_{\rm LST}=g^2_{\ndgt2}\theta_{\ndgt2}\ .\eea This means
that OM theory at a fixed energy $E$ and fixed effective coupling
$E\ell_{\rm OM}$ wrapped on a circle of small radius $ER<<1$ can
be described in terms of weakly coupled \gt2 theory.

\section{Summary and discussion\la{sec:dis}}

In this paper we have examined some implications of the fact that
a given bound state solution may be viewed either as an electric
deformation or as a magnetic deformation, and that the
corresponding near horizon limits lead to equivalent near horizon
regions. We propose that a new class of open D-brane theories,
the \od{q} theories, arise on D-branes with critical electric RR
fluxes, and that they are S-dual to NCOS or OD$p$ theories. We
have also argued that the \od{q} theories are related to NCYM
theories in limits of weak coupling involving NCYM solitons of
fixed tension. We then constructed duality webs, involving limits
of both strong coupling and large non-commutativity scale, and
showed that these require new theories with fixed spatial
non-commutativity, the D$q$-GT on D-branes and \gt{q} on
NS$5$-branes, that arise as perturbative descriptions in the
tensionless open and little string limits of NCOS and OD$p$
theory, respectively, at weak open string coupling or
OD$p$-coupling. An attempt to unify the duality diagrams for the
non-commutative theories assume the following shape:

{\scriptsize
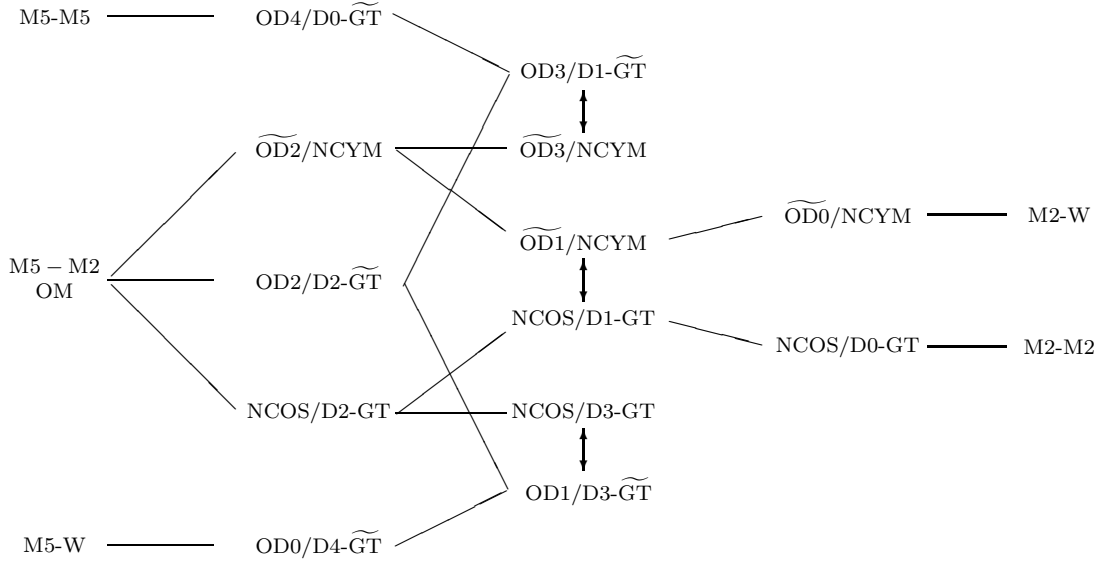
\begin{picture}(450,240)(0,-18)

\put(0,200){\makebox(0,0){M$5$-M$5$}}
\put(0,100){\makebox(0,0){$\begin{array}{c}{\rm M}5-{\rm
M2}\\{\rm OM}\end{array}$}} \put(0,0){\makebox(0,0){M$5$-W}}

\put(100,200){\makebox(0,0){OD$4$/\gt0}}
\put(100,150){\makebox(0,0){\od2/NCYM}}
\put(100,100){\makebox(0,0){OD$2$/\gt2}}
\put(100,50){\makebox(0,0){NCOS/D$2$-GT}}
\put(100,0){\makebox(0,0){OD$0$/\gt4}}

\put(200,150){\makebox(0,0){\od3/NCYM}}
\put(200,180){\makebox(0,0){OD$3$/\gt1}}
\put(200,115){\makebox(0,0){\od1/NCYM}}
\put(200,85){\makebox(0,0){NCOS/D$1$-GT}}
\put(202,20){\makebox(0,0){OD$1$/\gt3}}
\put(200,50){\makebox(0,0){NCOS/D$3$-GT}}

\put(300,125){\makebox(0,0){\od0/NCYM}}
\put(300,75){\makebox(0,0){NCOS/D$0$-GT}}

\put(380,125){\makebox(0,0){M$2$-W}}
\put(380,75){\makebox(0,0){M$2$-M$2$}}

\put(40,200){\makebox(0,0){\line(1,0){40}}}
\put(40,100){\makebox(0,0){\line(1,0){40}}}
\put(40,0){\makebox(0,0){\line(1,0){40}}}

\put(45,125){\makebox(0,0){\line(1,1){47}}}
\put(45,75){\makebox(0,0){\line(1,-1){47}}}

\put(150,190){\makebox(0,0){\line(2,-1){44}}}
\put(150,134){\makebox(0,0){\line(4,-3){41}}}

\put(152,60){\makebox(0,0){\line(1,-2){39}}}
\put(150,50){\makebox(0,0){\line(1,0){42}}}

\put(150,65){\makebox(0,0){\line(4,3){41}}}
\put(150,10){\makebox(0,0){\line(2,1){42}}}

\put(152,138){\makebox(0,0){\line(1,2){40}}}
\put(150,150){\makebox(0,0){\line(2,0){42}}}

\put(250,120){\makebox(0,0){\line(4,1){35}}}
\put(250,80){\makebox(0,0){\line(4,-1){35}}}

\put(345,125){\makebox(0,0){\line(1,0){30}}}
\put(345,75){\makebox(0,0){\line(1,0){30}}}

\put(200,164){\makebox(0,0){\vector(0,-1){8}\vector(0,1){8}}\makebox(7,-2)}
\put(200,100){\makebox(0,0){\vector(0,-1){8}\vector(0,1){8}}}
\put(200,36){\makebox(0,0){\vector(0,-1){8}\vector(0,1){8}}}

\end{picture}}

In this figure, in the first and fifth column we have theories
which originates from M theory, while the theories in column two
and four originate from type IIA and column three from type IIB
string theory. The horizontal and diagonal lines indicate
T-dualities between IIA/B theory or lifts/reductions between
IIA/M theory. The vertical lines imply S-dualities between
theories in IIB. We remark that they follow, in the usual way,
from the proposed modular invariance of M theory/OM theory
\cite{SchwarzBerman,us2}. For example, dropping on a transverse
circle from OM theory to OD$2$/\gt2, and then performing an
electric T-duality (i.e. a T-duality along a direction with
electric field) leads to OD$1$/\gt3. We can also reduce OM theory
on an electric circle to NCOS/D$2$-GT and then perform a
T-duality along a transverse direction, which leads to
NCOS/D$3$-GT. The S-duality between these two pairs (which are
internally related at small and large energies) now follows by
performing the modular transformation of the electric torus.
Similar arguments show the remaining two S-dualities in IIB from
the modular transformations of a magnetic torus and a
magnetic/electric torus.

The results in this paper points to the possibility of describing
D-brane dynamics using open D-branes rather than open strings in
weak coupling limits where the non-commutativity in the open
string sector becomes large. Let us end this discussion with some
speculations along these lines.

For NCYM (large spatial non-commutativity) this limit involves
NCYM solitons with fixed tension, which describe sub-critical
open D-branes as long as $\theta_{\rm YM}$ is finite. The angle
between the open D-brane and the D-brane on which it ends is then
non-zero, so there are no additional massless open string modes
in the spectrum.  The \od{q} theories, on the other hand, are by
definition assumed to have perturbative expansions in their
couplings $G^2_{\rm \ndod{q}}$ at fixed
$\theta_{\ndod{q}}=\theta_{\rm YM}g^2_{\rm YM}$. This formulation
is thus expected to involve critical open D-branes at zero angle.
At this critical angle the perturbative spectrum changes
abruptly, due to the new massless open string modes propagating
along the open D-brane. In particular, in the case of open
D-strings it is therefore reasonable to assume that the
interactions between the critical open D-strings are not the
ordinary fundamental self-interactions but rather interactions of
a new type, which are related to the other \od{q} interactions by
T-duality.

For NCOS (large temporal non-commutativity) this limit involves
tensionless open strings. Therefore it would be natural that in a
dual open D-brane picture, this limit would become a zero-slope
limit for open D-branes. In Section 5 we showed that at long
wave-length the resulting D$q$-GT's are weakly coupled indicating
a formulation in terms of some generalized gauge theory. For
$q=2,3$ these theories have two length scales; one, set by
$\theta_{Dq-GT}$, describing the non-commutative loop-space
structure of the theory; and a second much smaller one, set by
$g^2_{Dq-GT}$, determining the effective coupling of the theory.
This suggests a field theoretical realization with finite
(loop-space) non-commutativity while being in a region of weak
effective coupling at fixed energy. On the other hand, for $q=1$
the D$1$-GT may be considered at any length scale, so it should
be possible to remove the non-commutativity and go to an
undeformed theory in four dimensions, which should be some
extension of ordinary gauge theory.

We have also proposed S-duals of the D$3$-GT and six-dimensional
NCYM theory in the form of \gt3 and \gt1 on the type IIB
NS$5$-brane, and their T-duals \gt0, \gt2 and \gt4 on the type
IIA NS$5$-brane. These theories are tensionless limits of
non-commutative little string theory, which are strongly coupled
at long wave-length. However, the spatial non-commutativity
length scale can be taken to be much smaller, leaving room for a
phase where we expect that these theories have weakly coupled
formulations in terms of generalized gauge theories with spatial
(loop-space) non-commutativity on the IIB NS$5$-brane and tensor
analogs on the IIA NS$5$-brane. In the limit of vanishing
non-commutativity we expect these theories to be related to the
tensionless little string theories \cite{20,Hull}.

We also expect that the open D-brane theories could avoid some of
the non-perturbative issues that one would associate with
quantizing critical open M-theory membranes or finding other
field theoretical constructions of OM theory such as the
generalized tensor theories discussed above. Nevertheless we
found that the interactions of open D-brane based theories on the
IIA NS$5$ and D$4$-branes seem close to the OM theory
interactions, in the sense that the effective OM coupling can be
kept fixed at fixed energy in the limit of small radius where the
open D-brane theories become perturbative. Whether this brings
good news for attempts at understanding the OM theory, or bad
news for understanding open D$2$-branes, and other open
D$q$-branes remains to be seen.

Some of the results in this paper -- in particular on the \od{q}
theories in Section 4 -- overlaps with \cite{lu} which appeared
while we were preparing the manuscript.

\vspace{0.4cm}
\Large \textbf{Acknowledgements}
\vspace{0.2cm}
\normalsize

We thank E. Bergshoeff, M.~Cederwall, G.~Ferretti, U.~Gran, M.
~Nielsen, D.~Roest, J.P.~van der Schaar and E.~Sezgin
for interesting discussions. We are thankful to B.E.W.~Nilsson for discussions
and for valuable comments on an early version of the manuscript.
P.S. is thankful to the A\&M University and E.~Sezgin for hospitality during
the completion of this work. The work of P.S. is part of the research program
of ``Stichting voor Fundamenteel Onderzoek der Materie'' (FOM).
\appendix

\section{Conventions for S-duality, T-duality and lift to \\
11-dimensions}

In this appendix we give the conventions we have used for S-duality, T-duality
 and lift of type IIA solutions to 11-dimensions.

The S-duality rules for a type IIB solution (in the string frame) is given
 by (the axion is set to zero):

\bea \label{S}
d\tilde{s}^2&=& e^{-\phi}ds^2\ ,\quad e^{\tilde{\phi}}=e^{-\phi}\ ,
\nonumber\\
\tilde{B}&=&C_{2}\ ,\quad \tilde{C}_{2}=-B\ ,\\
\tilde{C}_{4}&=&C_{4}+B\wedge C_{2}\ .\nonumber
\eea

For uplifting of a type IIA solution to eleven dimensions we use the
following relations:

\begin{equation} {ds_{11}^2\over \ell_p^2}=e^{-2\phi/3}{ds_{IIA}^2\over
\alpha'}+e^{4\phi/3}({dx^{11}\over R}-{C_1\over \sqrt{\alpha'}})^2\ ,
\end{equation}
\begin{equation}
{A_3\over\ell_p^3}={C_3\over (\alpha')^{3/2}}+{dx^{11}\over R}\wedge {B_2\over
\alpha'}\ ,
\end{equation}
where $x^{11}$ has radius $R$ and $\ell_p$ is the eleven-dimensional
Planck length, which are given by $R=g\sqrt{\alpha'}$ and
$\ell_p=g^{1/3}\sqrt{\alpha'}$, where $g$ is the asymptotic value
of the dilaton.

For T-duality between IIA and IIB solutions we use the Buscher rules
\cite{bucher} (in the string frame)

\begin{eqnarray}\label{TNS}
\tilde{g}_{yy}&=&\frac{1}{g_{yy}}\ ,\quad \tilde{g}_{\mu\nu}=g_{\mu\nu}-
\frac{g_{\mu y}g_{\nu y}-B_{\mu y}B_{\nu y}}{g_{yy}}\ ,\nonumber\\
\tilde{B}_{\mu y}&=&\frac{g_{\mu y}}{g_{yy}}\ ,\quad \tilde{B}_{\mu\nu}=
B_{\mu\nu}-\frac{B_{\mu y}g_{\nu y}-g_{\mu y}B_{\nu y}}{g_{yy}}\ ,\\
\tilde{g}_{\mu y}&=&\frac{B_{\mu y}}{g_{yy}}\ , \quad e^{2\tilde{\phi}}=
\frac{e^{2\phi}}{g_{yy}}\ , \nonumber
\end{eqnarray}
where $y$ denotes the Killing coordinate with respect to which the
T-dualization is applied, while $\mu$,$\nu$ denote any coordinate direction
other then $y$. The RR $p$-form fields transforms as \cite{sei3}:

\begin{eqnarray}\label{TRR}
\tilde{C}_{(p)\mu\cdots \nu\rho y}&=&C_{(p-1)\mu\cdots \nu\rho}-
(p-1)\frac{C_{(p-1)[\mu\cdots\nu\mid y\mid}g_{\rho]y}}{g_{yy}}\ ,\nonumber\\
\tilde{C}_{(p)\mu\cdots \nu\rho\sigma}&=&C_{(p+1)\mu\cdots \nu\rho\sigma y}+
pC_{(p-1)[\mu\cdots\nu\rho}B_{\sigma]y}\\
& & +p(p-1)\frac{C_{(p-1)[\mu\cdots\nu\mid y\mid}B_{\rho\mid y\mid}
g_{\sigma] y}}{g_{yy}}\ .\nonumber
\end{eqnarray}

\end{document}